\documentclass{article}
\usepackage{epsfig}
\usepackage{amsmath}
\usepackage{amssymb}
\usepackage{amsthm}
\usepackage{xspace}
\usepackage{enumerate}
\usepackage{subcaption}

\newcommand\mm[1]{\ifmmode{#1}\else{\mbox{\(#1\)}}\fi}
\newcommand\Dgm{\mathrm{Dgm}}

\newsavebox{\smallProofsym}                            
\savebox{\smallProofsym}                               %
{
\begin{picture}(6,6)                                   %
\put(0,0){\framebox(6,6){}}                            %
\put(0,2){\framebox(4,4){}}                            %
\end{picture}                                          %
}                                                      %
\makeatletter
\long\def\@makecaption#1#2{%
  \vskip\abovecaptionskip
  \sbox\@tempboxa{\small #1: #2}%
  \ifdim \wd\@tempboxa >\hsize
    \small #1: #2\par
  \else
    \global \@minipagefalse
    \hb@xt@\hsize{\hfil\box\@tempboxa\hfil}%
  \fi
  \vskip\belowcaptionskip}
\makeatother

\newcommand\Gspace        {G}

\newcommand\Rspace        {\mm{{\mathbb R}}}

\newcommand\Vspace        {V}

\newcommand\Yspace        {\mm{{\mathbb Y}}}

\begin{document}

\title{Persistent homology analysis of brain artery trees}
\author{Paul Bendich%
        \thanks{Department of Mathematics, Duke University, Durham,
		North Carolina},
        J.S. Marron%
        \thanks{Department of Statistics and Operations Research, University
                of North Carolina, Chapel Hill, NC},
        Ezra Miller$^*\hspace{-1ex}$,
        Alex Pieloch$^*\hspace{-1ex}$,
        and
        Sean Skwerer$^\dagger$}
\date{23 November 2014}

\maketitle

\begin{abstract}
New representations of tree-structured data objects, using ideas from
topological data analysis, enable improved statistical analyses of a
population of brain artery trees.  A number of representations of each
data tree arise from persistence diagrams that quantify branching and
looping of vessels at multiple scales.  Novel approaches to the
statistical analysis, through various summaries of the persistence
diagrams, lead to heightened correlations with covariates such as age
and sex, relative to earlier analyses of this data set.  The
correlation with age continues to be significant even after
controlling for correlations from earlier significant summaries.
\end{abstract}

\section{Introduction}\label{sec:Intro}

In the modern era of large, complex data, statistical analysis is
particularly challenging when the sample points are not vectors but
rather objects with more intrinsic structure.  In the present case,
each data point is a tree, embedded in $3$-dimensional space, with
additional attributes such as thickness.  Background and additional
information concerning these data objects, which represent arteries in
human brains (isolated via magnetic resonance
imaging~\cite{aylward2002initialization}), occupy
Section~\ref{sec:Brain-artery-trees}.  Earlier analyses of this data
set have correlated certain features with age and produced hints of
sex effects~(Section~\ref{sub:Earlier-analyses}).

Topological data analysis (TDA), implemented here using persistent
homology, reveals anatomical insights unavailable from earlier
approaches to this data set (Section~\ref{sec:PH}).  In particular,
TDA shows age to be correlated with certain measures of how brain
arteries bend through space (Sections~\ref{sub:intuition}
and~\ref{sub:age-effects}).  This contrasts with a previous study
\cite{bullitt2005analyzing} that correlates age with total artery
length, and furthermore the TDA correlations are independent of that
earlier one (Section~\ref{sub:total-artery-length}).  TDA in our
context also finds stronger sex effects than the only other study
\cite{shen2013functional} to find any sex difference at all
(Section~\ref{sub:sex-effects}).

To quantify the bending of arteries through space, we rely on
persistent homology (Section~\ref{sec:TDA}), which we review from
scratch.  One of our methods records how the connectedness of the
subset of the vessels beneath a given horizontal plane changes as the
plane rises from below the brain to above it
(Section~\ref{sub:height}).  Another of our methods records the
evolution of independent loops contained in the
$\varepsilon$-neighborhood of the tree as $\varepsilon$ increases from
$0$ to~$\infty$ (Section~\ref{sub:thickening-and-loops}).  In either
method, for each tree the topological information is recorded in a
persistence diagram, which is a finite set of points above the main
diagonal in the positive quadrant of the Cartesian plane.  These
diagrams are turned into feature vectors in a variety of ways,
resulting in several statistical analyses, detailed in
Section~\ref{sec:Detailed}.

This work fits into the framework of Object Oriented Data Analysis
(OODA, \cite{wang2007object}, and also
\cite{marron2014overview} for a current
overview).  In particular, the concept of \emph{data object} here
serves as a platform for interdisciplinary discussion of a type that
lies at the core of complicated data analyses.  The transition from
$3$-dimensional tubular image to persistence diagram to feature
vector, with only the latter being amenable to standard linear
statistics, demonstrates the power of the notion of data object to
incorporate data \emph{representation}, in contrast to the concept of
\emph{experimental unit} in 
\cite{lu2014object}.

\section{Brain artery trees}\label{sec:Brain-artery-trees}

\begin{figure}
\begin{center}
\includegraphics[scale=0.5]{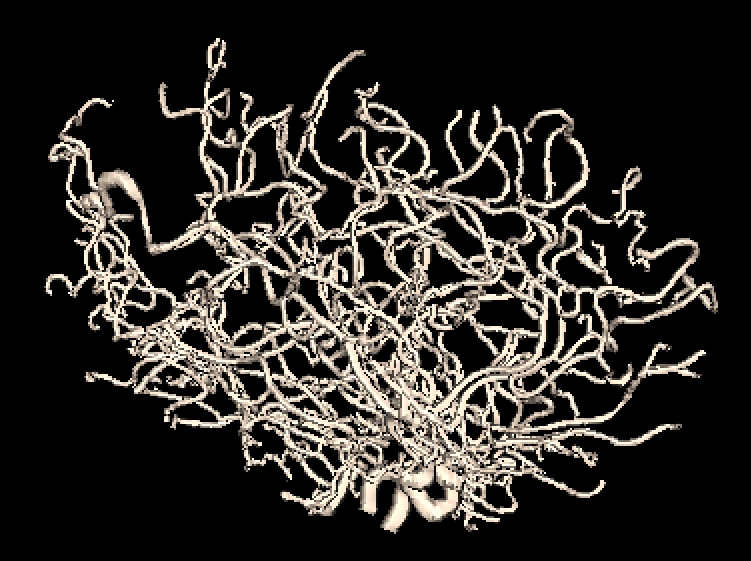}
\vspace{-1ex}
\end{center}
\caption{\label{fig:DataObject}%
Tree of arteries from the brain of one person, showing one data
object.  Thickest arteries appear near the bottom.  Arteries bend,
twist, and branch through three dimensions, which results in
meaningful aspects of the data being captured by persistent homology
representations.}
\end{figure}

Each data point in this study is the tree of arteries in the brain of
one person, collected by tube-tracking vessel segmentation from
3-dimensional Magnetic Resonance Angiography (MRA) images followed by
a combination of automatic and manual assembly into trees.  Aylward
and Bullitt \cite{aylward2002initialization} and Aydin et
al. \cite{aydin2009principal} describe and discuss this
process.  A visual (2-dimensional) rendering of one such reconstructed
tree is shown in Figure~\ref{fig:DataObject}.  The full data set
consists of $n = 98$ such trees from people whose ages range from 18
to 72 years old.
While the long-term goal of study is to develop methods for exploring
stroke tendency, and perhaps to develop diagnostics for brain cancer
based on vasculature, pathological cases were deliberately excluded
from the Handle dataset, with the idea of first understanding
variation in the population of non-pathological cases.  The only
remaining interesting covariates are age and sex; understanding
correlations with these is the central~goal~of~this~study.

\subsection{Earlier analyses}\label{sub:Earlier-analyses}

Among the earlier analyses of this and closely related data sets, the
first was by Bullitt, et al. in \cite{bullitt2005analyzing}, who
studied simple summaries of each data tree, such as overall branch
length and also average branch thickness, which were both seen to be
significantly correlated with age.  A clear avenue for refinement of
that approach is to make better use of the large amount of additional
information available in this rich data set, such as the tree
topologies, and also the multiple individual branch locations,
structures, widths, and so on.  Early approaches to this, such as 
\cite{wang2007object} or \cite{aydin2009principal}, chose to focus solely on the
combinatorics of the branching structure, ignoring other aspects such
as thickness or the geometry of the 3-dimensional embedding.  The
latter paper found statistically significant age effects.  These age
effects were studied more deeply using the notion of tree smoothing
developed by Wang et al. in \cite{wang2012nonparametric}.

A much different approach to this data set was taken by Shen et
al. in \cite{shen2013functional}, based on representations of
planar binary trees via Dyck paths, an early appearance of which came
in probability theory of branching processes \cite{harris1952first}.
The bijection represents each planar binary tree as a function,
allowing application of standard asymptotic methods when trees are
viewed as random objects.  Adaption to the brain artery tree dataset
had the goal of making available the large array of methods available
for Functional Data Analysis (FDA), where the data objects are curves
such as graphs of univariate functions; see \cite{ramsay2002applied,ramsay2006functional}.  Dyck path
analysis of the brain tree data found more significant correlation
with age as well as the first indication of a significant
sex~effect.

A drawback of the above approaches to tree data analysis is that they
require 2-dimensional embedding of the given 3-dimensional tree
structure, as noted in Section~2.1 of
\cite{aydin2009principal} and Section~2.1 of Shen et
al. \cite{shen2013functional}.  For each non-leaf node, a
choice must be made as to which child node goes on the left and which
goes on the right (if the tree is not binary, then an ordering of the
node's children is required).  While ad hoc methods were used to
reasonable effect in those papers, it is natural to suspect that they
result in loss of statistical efficiency.  This issue can be seen as
an instance of the \emph{correspondence problem}: planar embedding
necessarily violates any consistent, anatomically meaningful
assignment of $3$-dimensional features across objects in this dataset.

An approach to overcoming this problem is based on the concept of
phylogenetic tree from evolutionary biology; see \cite{holmes1999phylogenies} for a good introduction to this
area, and \cite{billera2001geometry} for an
insightful mathematical treatment.  A major challenge in applying this
idea to a set of brain artery trees is that phylogenetic trees require
a fixed underlying set of leaves, while brain artery trees have leaves
that appear where the vessel thickness passes below the imaging
resolution of MRA, locations of which vary across cases.  One way to
resolve this problem was implemented by Skwerer et
al. \cite{skwerer2013tree}, who used additional cortical
surface information plus a correspondence technique to produce a
common set of landmarks, which became the leaves.  That paper went on
to find statistically significant age and gender effects, some of
which were stronger than those previously found.

For additional treatments of tree-structured data objects that did not
analyze this data set, Feragen et
al. \cite{feragen2011geometries} developed an approach that
avoids both the planar embedding and fixed-leaf-set problems, and Nye
\cite{nye2011TreePCA} invented an analogue of principal
component analysis for phylogenetic trees.  Hotz et
al. \cite{hotz2012sticky}, followed by Barden et al. \cite{BardenLeOwen2013, BardenLeOwen2014} have investigated
surprising non-standard central limit theory in phylogenetic tree
spaces.

\section{Persistent homology analysis of brain arteries}\label{sec:PH}

The present investigation takes a completely different approach, using
topological data analysis (TDA); see
\cite{Edelsbrunner2010} for a good introduction to that set of
ideas.  We postpone to Sections~\ref{sec:TDA} and~\ref{sec:Detailed} a
review of precise definitions of key terms from TDA and the detailed
extraction of useful features for statistical analysis.  In this
section, we confine ourselves to a loose, intuitive description of
these features, and we describe the striking age and sex effects found
using this new feature set.  We also demonstrate that these effects
are independent of coarser geometric measures, such as total artery
length or average branch thickness, used in the earliest analyses.

\subsection{Intuition}\label{sub:intuition}

\begin{figure}
\begin{center}
\includegraphics[scale=0.2]{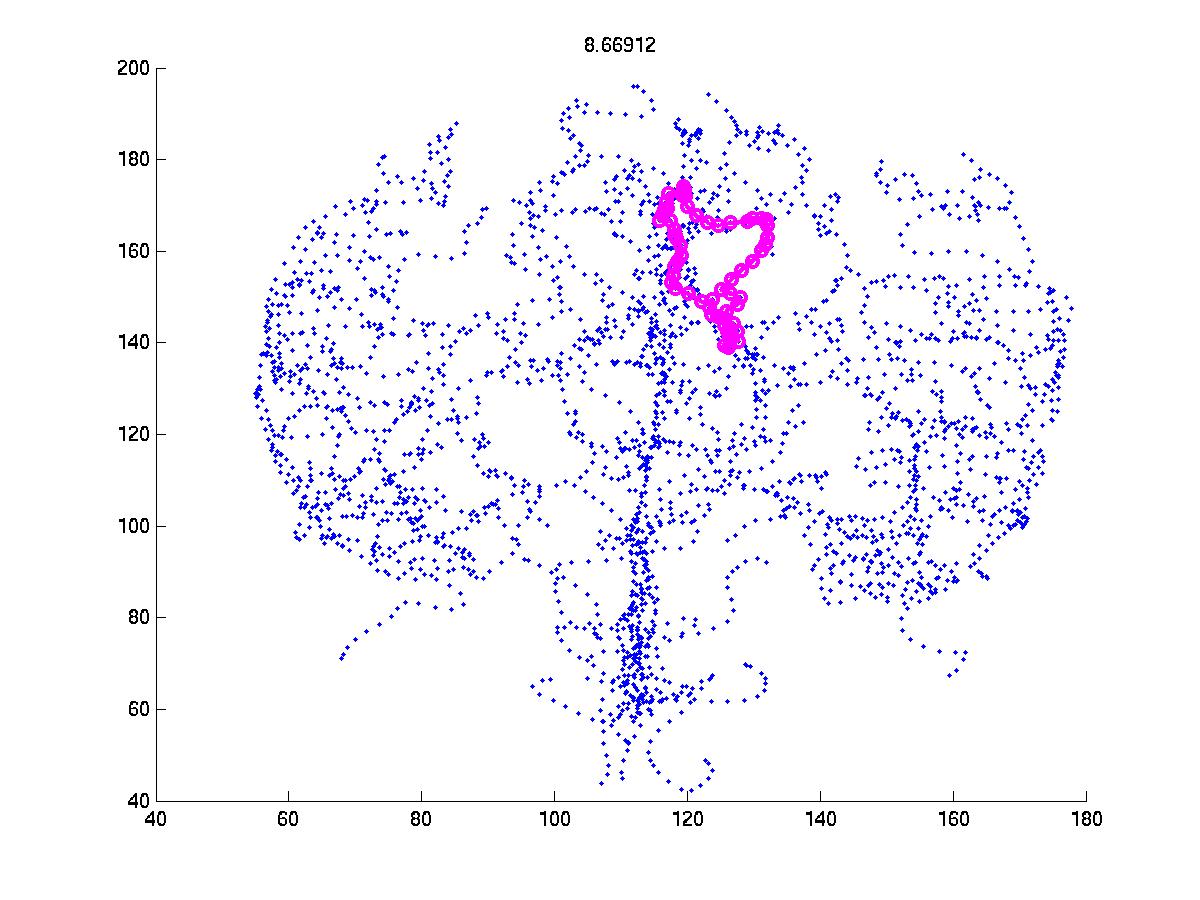}
\vspace{-3ex}
\end{center}
\caption{\label{fig:Case1Gen12}%
A MATLAB rendering of the brain artery tree of Patient 1.  Highlighted
in pink is one of the loops formed by thickening the artery tree
within the brain.  Also found are some of the loops and bends made by
the artery tree within the 3-dimensional geometry of the brain.}
\end{figure}

The methodology developed here provides a direct and quantitative
description, in the form of numerical features usable for statistical
analysis, of the way arterial structure occupies space within the
$3$-dimensional geometry of the brain.  Although a full description is
deferred to the next section, we illustrate some of what this means
here, with the aid of the tree in Figure~\ref{fig:Case1Gen12}.

First, looking near the bottom of Figure~\ref{fig:Case1Gen12}, notice
a large $S$-shaped bend in the arterial structure.  Bends such as
these, and other much tighter bends, occur throughout the tree.  The
technique of zero-dimensional persistent homology locates these bends
and measures their sizes.  TDA summarizes these sizes as a sequence
$p_1 > p_2 > \cdots $ of non-negative numbers, where $p_i$ is the size
of the $i$-th largest bend in a particular brain.

For a different flavor of geometry, imagine gradually thickening each
artery so that the tree begins to fill the $3$-dimensional space
containing it.  Loops start to form (for example, the one outlined in
pink in the figure) and then eventually fill in.  The time between
when a loop forms and when it fills in is called its persistence.  The
technique of one-dimensional persistent homology locates these loops
and measures their persistences, resulting in another sequence $q_1 >
q_2 > \cdots$ of non-negative numbers.

Rigorous definitions of the above terms are given in the next two
sections.  For the rest of this section, we suppress such details and
focus only on the analysis of features derived from persistence.

\subsection{Age effects}\label{sub:age-effects}

\begin{figure}[h]
\begin{center}
\includegraphics[scale=0.65]{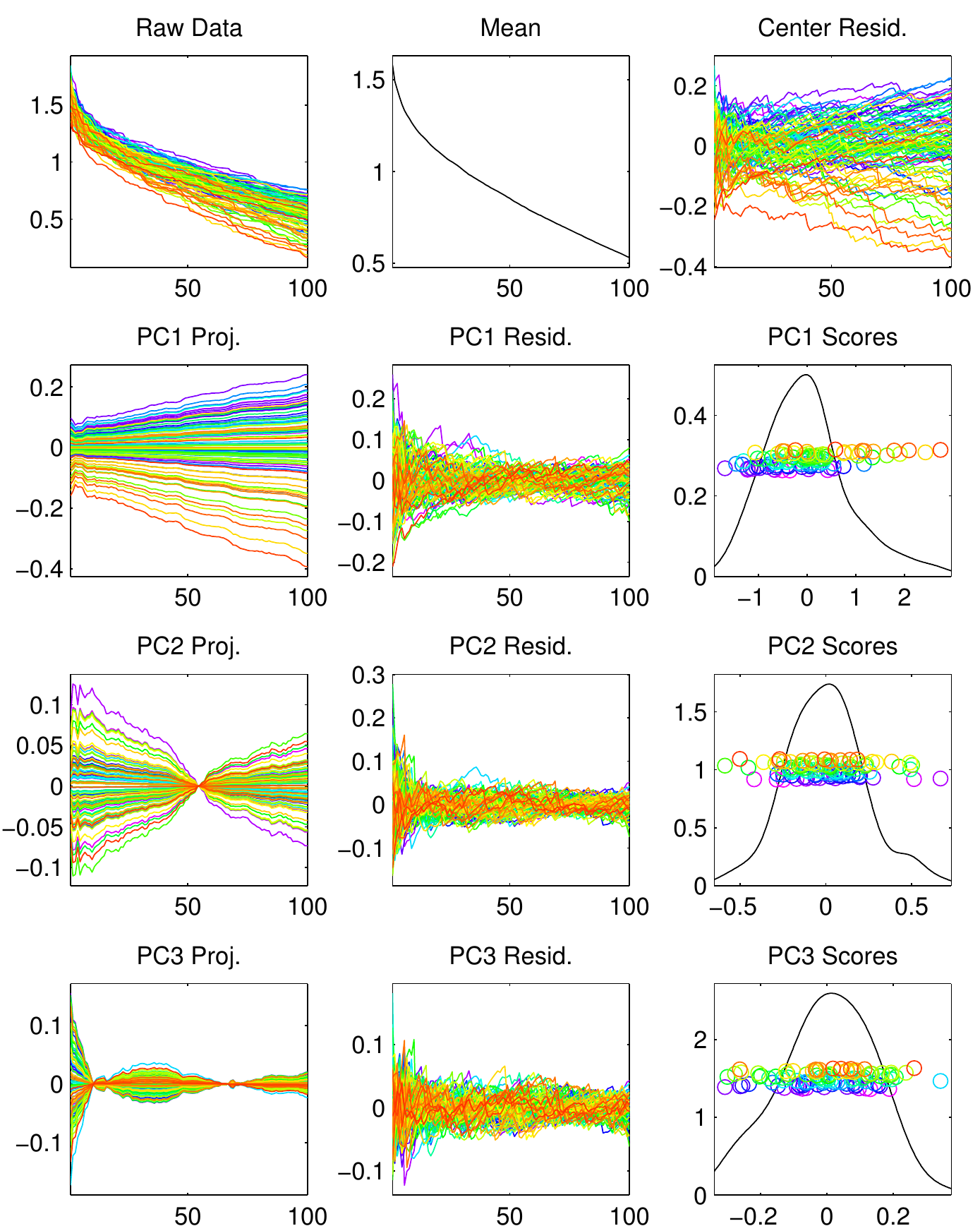}
\vspace{-2ex}
\end{center}
\caption{\label{fig:CurvDat}%
PCA of vector representations.  Raw data, mean and mean residuals are
in the top row.  Other rows show loadings and scores for the first 3
PCs, i.e.~modes of variation.  Rainbow colors indicate age.
Correlation of PC1 and age is apparent, with warmer colors generally
at the bottom and cooler colors generally at the~top.}
\end{figure}

Figure~\ref{fig:CurvDat} depicts a first population level view of the
sets of numbers $p_1, \ldots, p_{100}$ across the entire data set of
$n = 98$ brain trees.  Two out of three panels in each row contain a
set of $n = 98$ overlaid curves, each of which is a parallel
coordinates plot (see \cite{inselberg1997multidimensional} for an overview of this
graphical device): the coordinates of each data vector are plotted as
heights on the vertical axis as a function of the index, in this case
$i = 1, \ldots, 100$.  Color denotes age via a rainbow scheme starting
with magenta for the youngest~(19), ranging smoothly through blue,
green, yellow to red for the oldest~(79).  The upper left panel shows
the data curves, which already exhibit potential age structure in this
sample, with the younger people tending to appear near the top, and
the older people near the bottom.

As elegantly illustrated in \cite{ramsay2002applied,ramsay2006functional}, Principal
Component Analysis (PCA) can reveal deeper structure in a sample of
curves; see \cite{jolliffe2005PCA} for an introduction
to that method in general.  PCA starts at the center of the data: the
mean shown as a curve in the top center panel.  Variation about the
mean is studied through the mean residuals, which are the data curves
minus the mean, shown in the top right panel.  PCA next investigates
\emph{modes of variation} by finding orthogonal projections in the
curve space that represent maximal amounts of variation.  Projections
corresponding to the first PC are shown in the left panel of the
second row, which gives a more focused impression of younger people on
the top, and older people near the bottom.  In PCA terminology, this
is called a \emph{loadings plot}, because each curve is merely a
multiple of the first eigenvector, whose entries are called loadings.
The shape of the curve gives insight concerning this component (mode
of variation).  In this case the variation is all values moving in
unison, being either large or small together.  The center plot in the
second row shows the remaining variation, after subtracting the first
component from the centered residuals.  Careful study of the vertical
axes shows there is much less variation present.  The second row right
panel shows the PC1 scores as horizontal coordinates of points (with
age ordering for the vertical coordinates) with corresponding rainbow
colors, which are the coefficients of the projections shown in the
left panel.  These show a clear correlation between age and PC1.

The second PC explains as much variation in the data as possible among
directions orthogonal to PC1.  This shows the second major mode of
variation in the data, which differs in how quickly the values taper
off.  Similarly for PC3 (orthogonal to both earlier directions) in the
bottom row.  Note that neither PC2 nor PC3 seems to have much visual
connection with age, suggesting that most age effects have been
captured by PC1.

\begin{figure}[h]
\begin{center}
\includegraphics[scale=0.45]{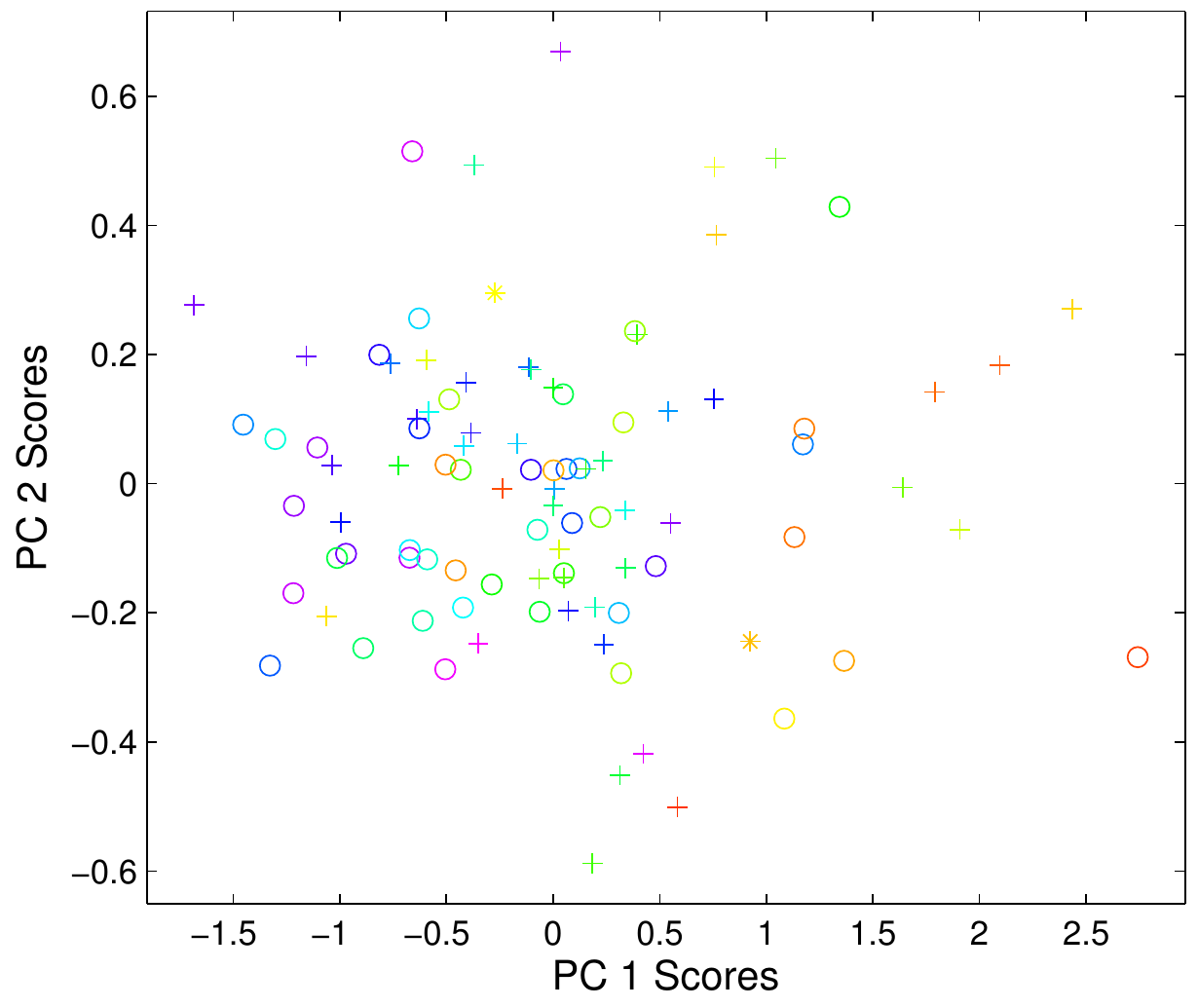}
\vspace{-2ex}
\end{center}
\caption{\label{fig:PC1PC2}%
Scatterplot of PC1 vs.~PC2.  Shows joint distribution of scores.  Main
lesson is PC1 appears strongly correlated with age, but not PC2 does not.}
\end{figure}
Figure~\ref{fig:PC1PC2} depicts an alternate PCA view of the data.
This is a \emph{scores scatterplot}: the scores are the coefficients
of the projections.  Here each symbol represents a person (same
rainbow coding for age color scheme as in Figure~\ref{fig:CurvDat}),
with the PC1 score plotted on the vertical axis, and the PC2 score on
the horizontal.  This scatterplot is the most variable two-dimensional
projection of the data and thus is generally useful for understanding
relationships between data objects.  In the current case, again the
rainbow color suggests an age gradient in the horizontal (PC1)
direction.  Figure~\ref{fig:PC1PC2} also allows study of sex using
symbols, with females represented by circles and males by plus signs.
No apparent visual gender differences can be seen here, but it is
important to keep in mind this is only a two-dimensional view of a
100-dimensional data space.  This issue is explored more deeply at the
end of this section.

\begin{figure}[h]
\begin{center}
\includegraphics[scale=0.45]{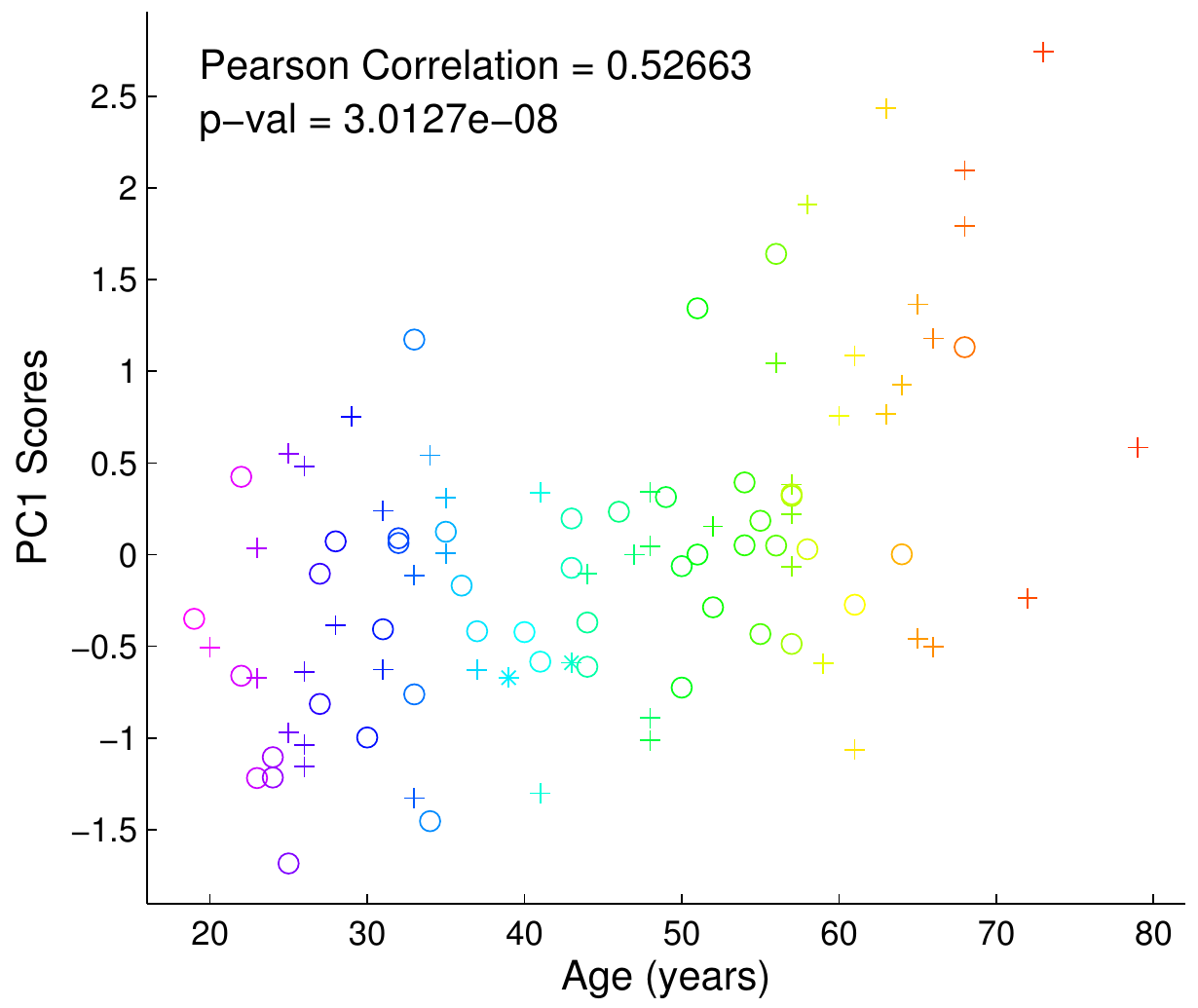}
\vspace{-2ex}
\end{center}
\caption{\label{fig:pcAge0}%
PC1 vs.~age for the zero-dimensional topological features verifies
strong correlation of PC1 with age.}
\end{figure}

A more direct study of the correlation between PC1 and age appears in
Figure~\ref{fig:pcAge0}, which plots the PC1 score as a function of
age.  The correlation is visually present, and the Pearson correlation
of $\rho = 0.53$ reflects it.  A simple Gaussian-based hypothesis test
against the null hypothesis of no correlation shows a strongly
significant $p$-value $< 10^{-7}$.

\begin{figure}
\begin{center}
\includegraphics[scale=0.45]{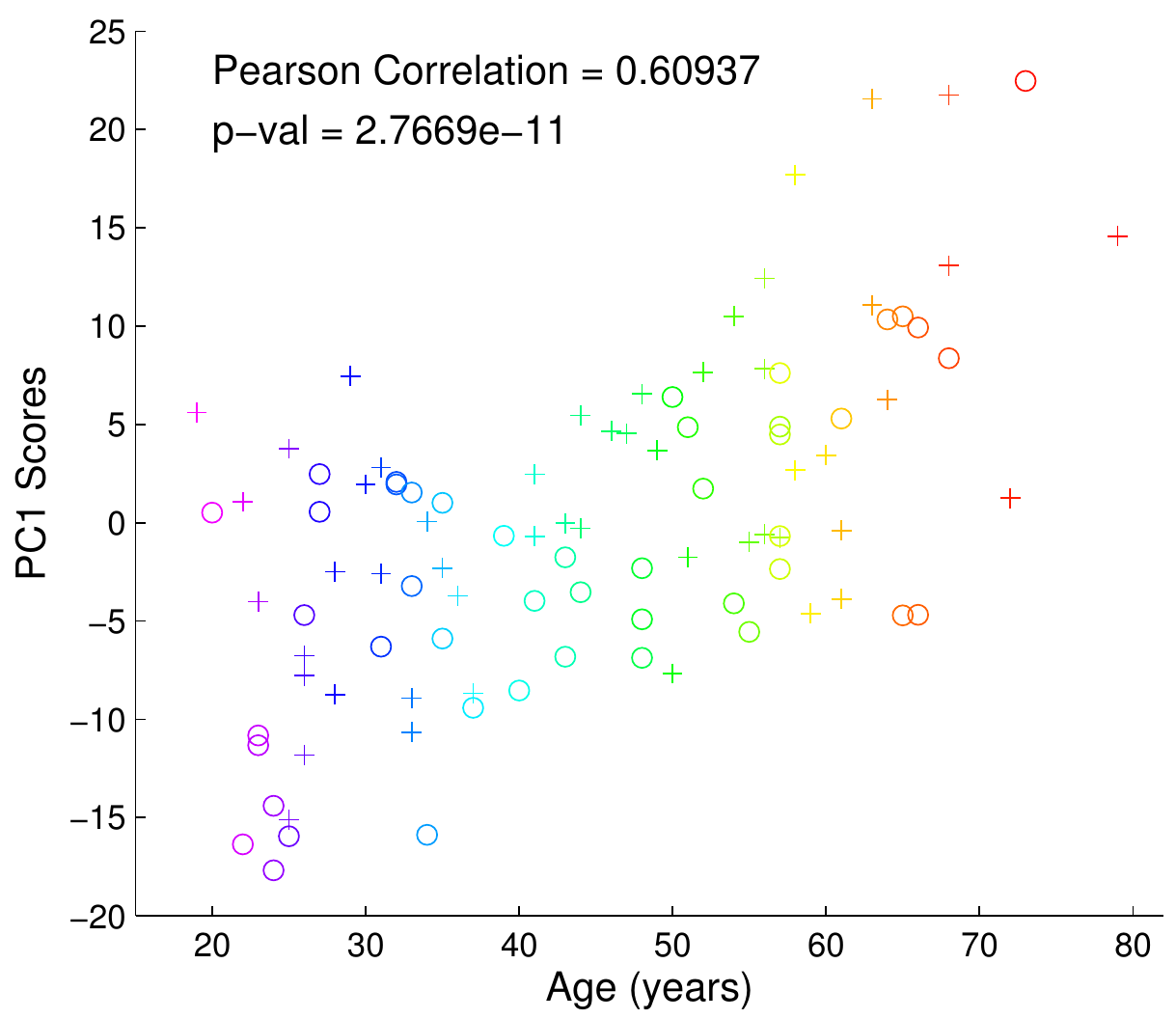}
\vspace{-2ex}
\end{center}
\caption{\label{fig:pcAge1}%
PC1 vs.~age for the one-dimensional topological features exhibits even
stronger correlation.}
\end{figure}

The same analysis can be repeated for the loop-persistence-based
numbers $q_1, \ldots q_{100}$.  Omitting the lengthy development
above, the key result can be seen in Figure \ref{fig:pcAge1}, namely
that the correlation between PC1 and age is even stronger for this
feature set: $\rho = 0.61$ with a $p$-value $<10^{-10}$.

\subsection{Total artery length}\label{sub:total-artery-length}

Early exploration of these data by Bullitt et
al. \cite{bullitt2005analyzing} demonstrated that younger
patients tend to have longer total artery length $L$, a statement that
might lead (as it led us) to justifiable skepticism about the novelty
of our findings.  More precisely, we quantify the sizes of artery
bends (and loops) at different scales, and these sizes could plausibly
be controlled by the total artery length of the tree: speaking
loosely, an attempt to jam a longer tree into the same skull would
naturally lead to more bending behavior.

To ensure we were not merely applying a complicated TDA machine to
detect a simple geometric phenomenon, we performed a more
sophisticated analysis.  For each $i$, linear regression between the
variables $p_i$ and~$L$ yields a residual $\hat{p_i}$.  Replacing
$p_i$ by~$\hat{p_i}$ in the analysis from
Section~\ref{sub:age-effects} results in an equally strong Pearson
correlation of $\rho = 0.52$, with a $p$-value on the order
of~$10^{-8}$.

Geometrically motivated methods to control for effects of total artery
length yield similarly negligible increases or decreases in Pearson
correlation and $p$-value.  These methods simply divide the
numbers~$p_i$ by (i)~$L$ or (ii)~$\sqrt L$ or (iii)~$\sqrt[3] L$
before running the analysis in Section~\ref{sub:age-effects}.  The
exponents on~$L$ correspond to physical models where vessel length
(i)~scales according to total linear skull size, (ii)~has constant
flux (i.e.~number of arteries passing) through each unit of
cross-sectional area, or (iii)~remains constant per unit volume.

The strength and significance of correlation after controlling for
total length breaks new ground in the analysis of the brain artery
data.  In particular, the persistent homology analysis here is
sensitive to genuine multi-scale geometrical structure in the arterial
systems, and does not simply reflect coarse size aspects of the data.

Controlling for total length in the one-dimensional persistence
analysis from Section~\ref{sub:age-effects} yields decidedly weaker
(but still non-negligible) age correlation: replacing the~$q_i$
features with their residuals~$\hat{q_i}$, after running a linear
regression between each $q_i$ and~$L$, results in Pearson correlation
$\rho = 0.35$.

\subsection{Sex effects}\label{sub:sex-effects}

Also of interest is an investigation of potential differences in brain
artery structure between the sexes.  Figure~\ref{fig:PC1PC2} provides
a preliminary view: male cases are indicated with a plus sign, and
females are shown as circles.  As noted in
Section~\ref{sub:age-effects}, meaningful sex difference is not
apparent in this plot, perhaps because PC1 seems to be driven more by
the independent age difference.  However, in high-dimensional data
analysis, simple visualization of the first few PC components is
frequently revealed to be an inadequate method of understanding all
important aspects of such data, because it is driven entirely by
variation, which can be a different goal.

\begin{figure}
\begin{center}
\includegraphics[scale=0.5]{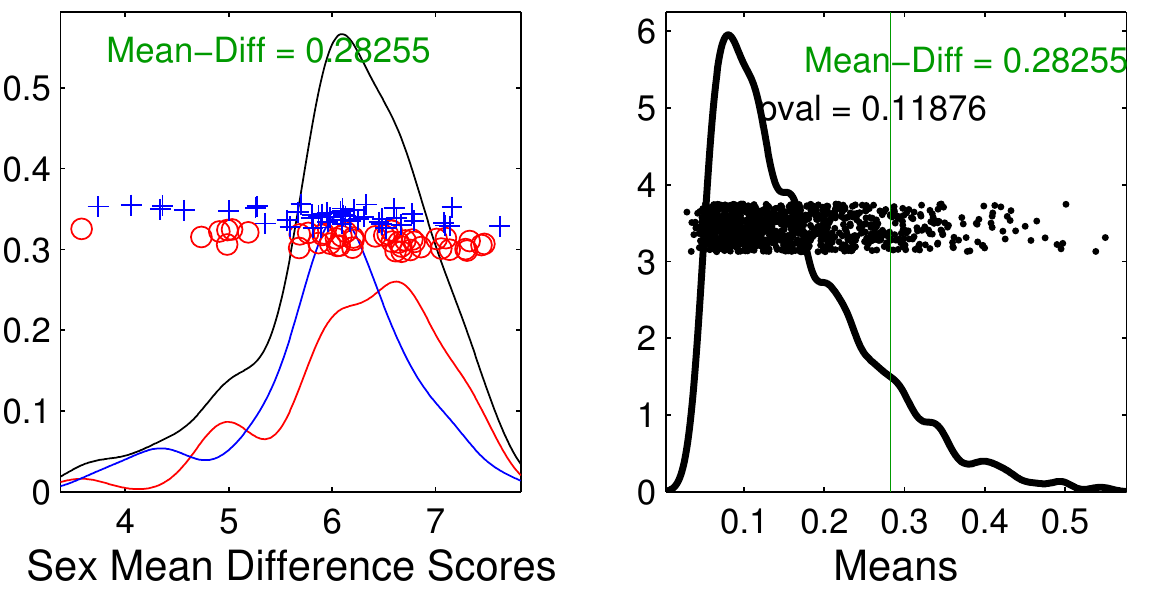}
\end{center}
\caption{\label{fig:DPP}%
Illustration of DiProPerm results on the zero-dimensional persistence
features.  The left panel shows the result of projecting the data onto
the direction vector determined by the means, suggesting some
difference.  The results of the permutation test are shown on the
right, with the proportion of simulated differences that are bigger
than that observed in the original data giving an empirical $p$-value.}
\end{figure}

A way to focus on desired effects in higher dimension is to calculate
the arithmetic mean of the vectors $(p_1, \ldots, p_{100})$
corresponding to male subjects, to do the same for the female
subjects, and then compute the Euclidean distance between these means
in $\mathbb{R}^{100}$.  The size of this mean difference alone does
not tell much as a raw number, but a simple permutation test on the
mean-difference statistic reveals more: randomly reassign the $98$
vectors into two groups of equal size, compute the difference between
the means of the two groups, and repeat this procedure $1000$ times.
This method has been called DiProPerm in \cite{wei2013direction}, and is illustrated in
Figure~\ref{fig:DPP}.

In our test, $119$ of the reassignments led to a larger
mean-difference than the original male--female split, giving an
estimated $p$-value of $0.1$, which is not impressive.  However, we
then repeated the entire procedure for the loop-vectors $(q_1, \ldots,
q_{100})$, and found a more compelling $p$-value of $0.032$.

In Section~\ref{sec:Detailed}, we demonstrate that a more thorough
analysis of feature selection results in even lower $p$-values for sex
difference.  These results are stronger than those in \cite{shen2013functional}, which is the only other study to
find a statistically significant sex difference in this data set.

\section{Topological data analysis methods}\label{sec:TDA}

We now give a more thorough discussion of topological data analysis,
in particular the zero- and one-dimensional persistence diagrams, from
which the features $p_i$ and $q_i$ in Section~\ref{sec:PH} are
extracted.  Put briefly, a persistence diagram provides a compact
two-dimensional record of the geometric and topological changes that
occur as an object in space is built in stages.

A rigorous description of persistence diagrams in their most general
context requires background from algebraic topology.  Fortunately, the
applications in this paper involve the simplest type of persistence
diagram, which tracks the appearance and disappearance of connected
components in a filtered graph, as well as a slightly more complicated
diagram, which tracks the formation and destruction of loops in a
thickening object.  This section contains a fully rigorous description
of the first type of diagram, a broadly intuitive description of the
second, and the details of how our initial artery tree data objects
produce both types.  Interpretation of these diagrams across a
population requires the statistical analysis outlined in
Section~\ref{sec:Detailed}, but the current section demonstrates, via
a few examples, the major features they are meant to isolate.

For more information on persistence diagrams in general, see \cite{Edelsbrunner2010} or
\cite{Carlsson2009}.  For a fully
detailed, rigorous exposition of what we outline here, see \cite{Chazal2009b}.

\subsection{Height functions and connected components}\label{sub:height}

\subsubsection{Graphs and critical values}\label{sub:graphs}

\begin{figure}
\begin{center}
\includegraphics[scale=0.35]{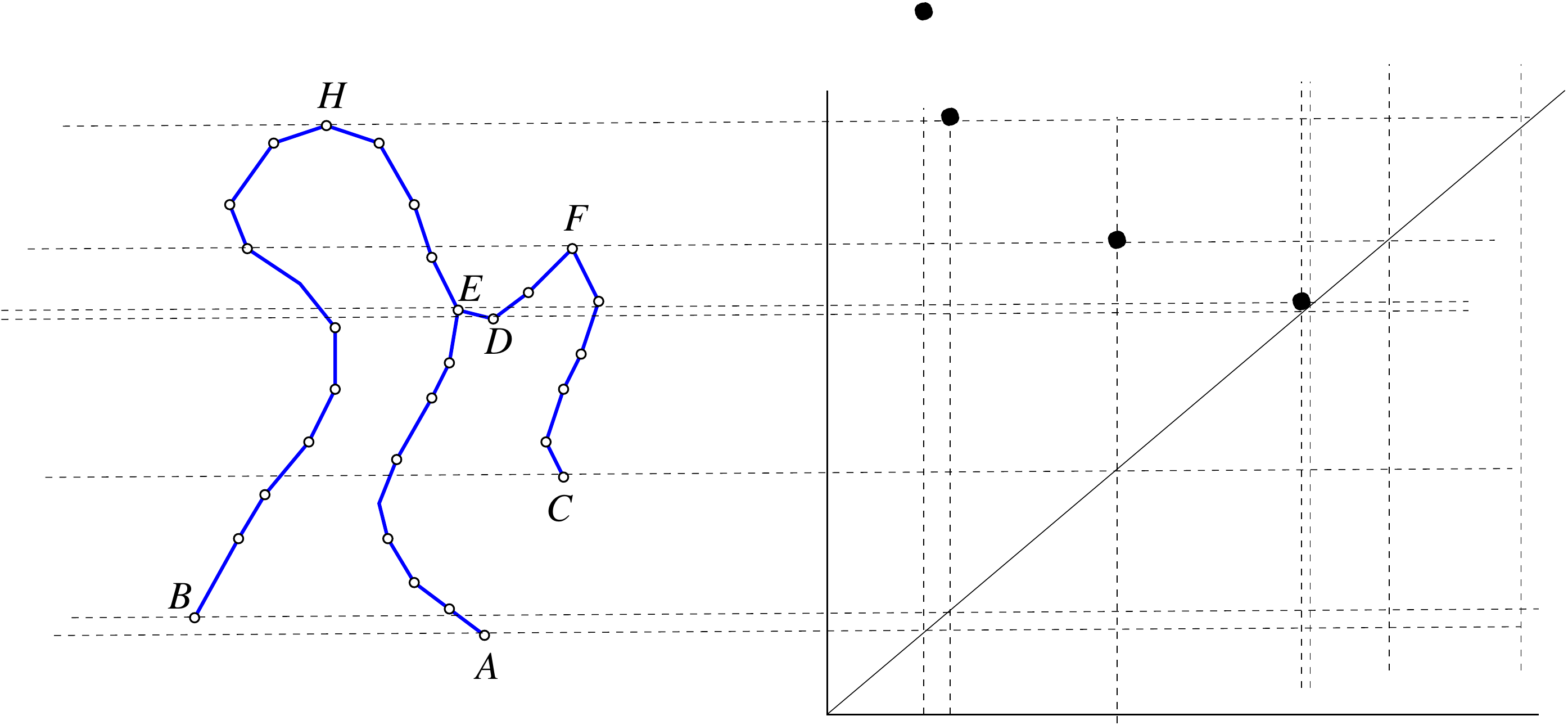}
\end{center}
\caption{\label{fig:TDE}%
On the left, a graph $G$.  The function $h$ measures height in the
vertical direction, and the persistence diagram $\Dgm_0(h)$ is shown on
the right.  The coordinates of the dots are, reading from right to
left, $(h(A), \infty), (h(B),h(H)), (h(C),h(F)),$ and $(h(D),h(E))$.}
\end{figure}

Let $\Gspace$ be a graph \cite{west1996graphtheory}, which loosely
speaking is a set~$V$ of vertices with specified pairs from~$V$ joined
by edges.  Fix a real-valued function $h: \Vspace \to \Rspace$.  For
simplicity of exposition, assume $h(v) = h(w)$ only if $v = w$.  As a
working example, let $\Gspace$ be the graph embedded in the plane as
shown on the left side of Figure~\ref{fig:TDE}, and let $h(v)$ be the
height of vertex $v$ as measured in the vertical direction.
Extend~$h$ to a function on the edge set by setting $h((v,w)) =
\max(h(v),h(w))$ for each edge~$(v,w)$~of~$\Gspace$.

The persistence diagram $\Dgm_0(h)$ takes $\Gspace$ and~$h$ as input
and returns as output a multi-scale summary of the component evolution
of the threshold sets of~$h$.  This output is robust with respect to
small perturbations of~$h$ \cite[Section~3]{CohenSteiner2007}.
We now explain in more detail what this means.

For each real number~$r$, define $\Gspace(r)$ to be the full subgraph
on the vertices with $h$-value at most~$r$.  For example, in
Figure~\ref{fig:TDE}, $\Gspace(r)$ is empty if $r < h(A)$ and
$\Gspace(r) = \Gspace$ whenever $r \geq h(H)$.  The graph $\Gspace$
itself consists of only one connected component, but we are far more
interested in what happens for values of~$r$ between $h(A)$
and~$h(H)$.

To be precise, label the vertices $v_1, \ldots, v_N$ by ascending
order of $h$-value, choose real numbers $r_i$ such that $h(v_i) < r_i
< h(v_{i+1})$, and set $\Gspace(i) = \Gspace(r_i)$.  Define the
\emph{lower link} $L(i)$ of the vertex $v_i$ to be the set of vertices
adjacent to~$v_i$ that have lower $h$-value than $v_i$ does.
Persistent homology records how the number $\beta_0(i)$ of connected
components of~$\Gspace(i)$ changes as $i$ increases.

\begin{figure}[ht]
\begin{center}
\hfill
\includegraphics[angle=270,origin=c,scale=0.35]{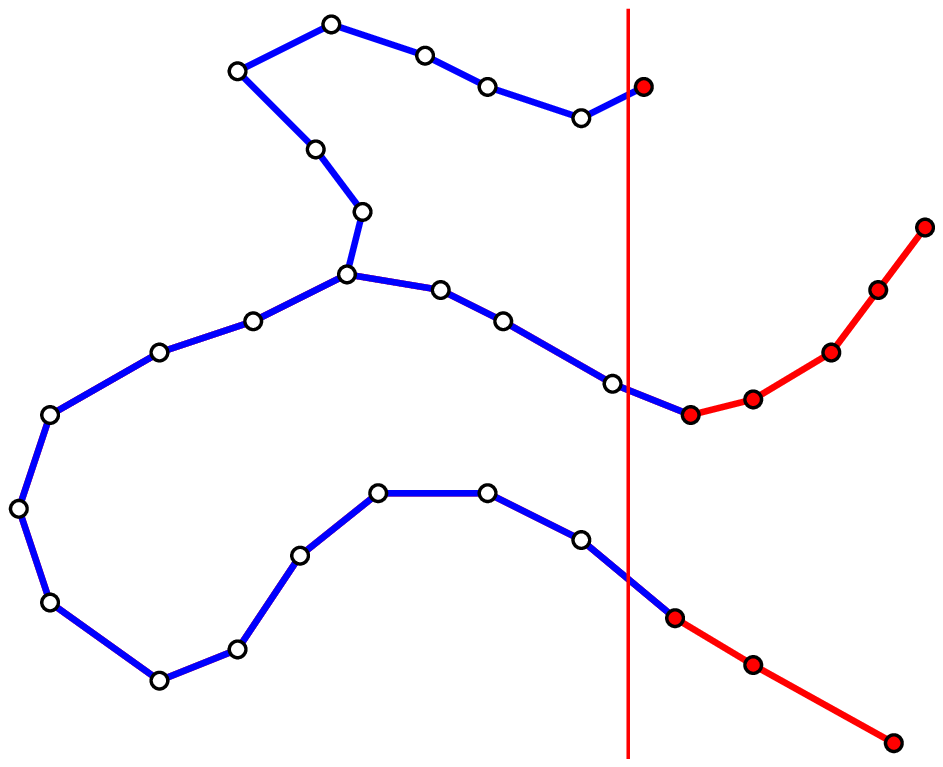}\hfill
\includegraphics[angle=270,origin=c,scale=0.35]{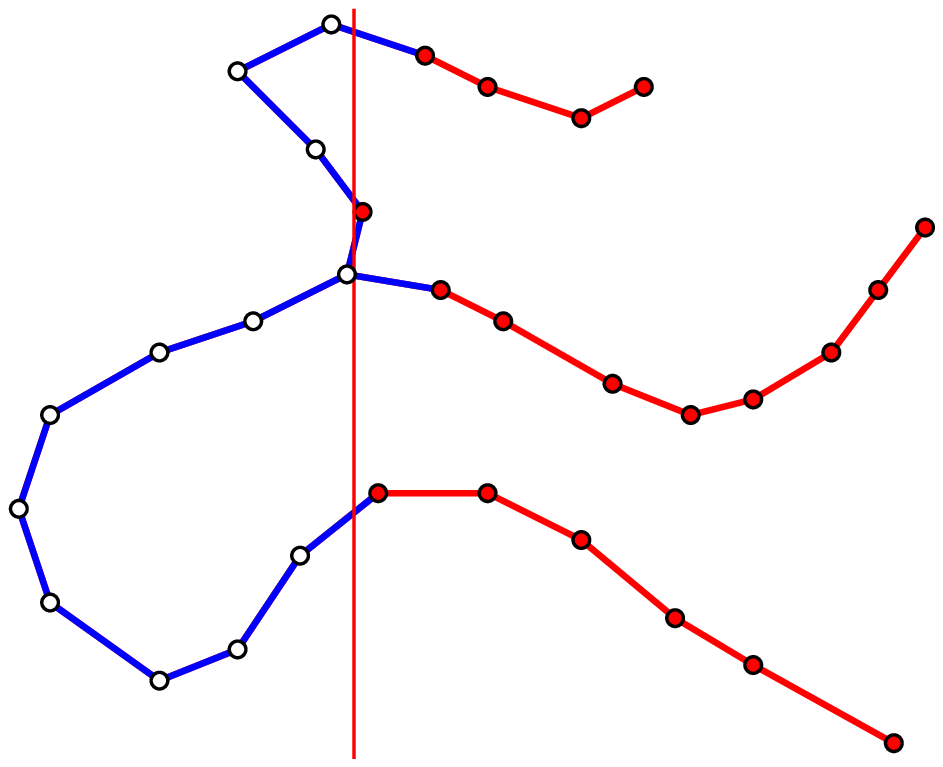}\hfill
\includegraphics[angle=270,origin=c,scale=0.35]{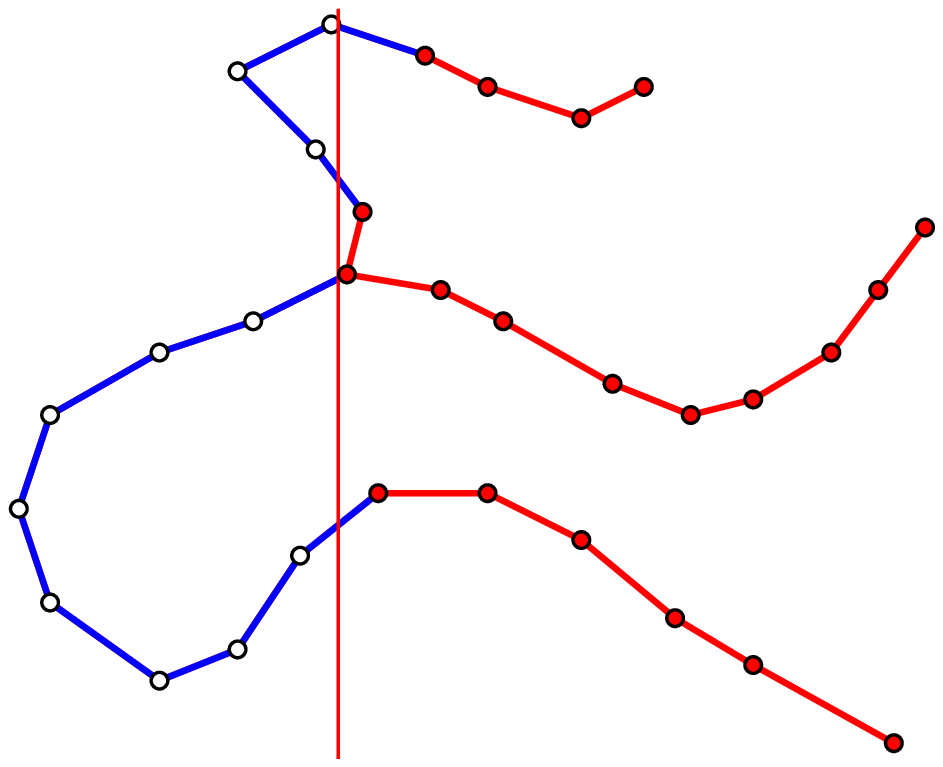}\hfill
\includegraphics[angle=270,origin=c,scale=0.35]{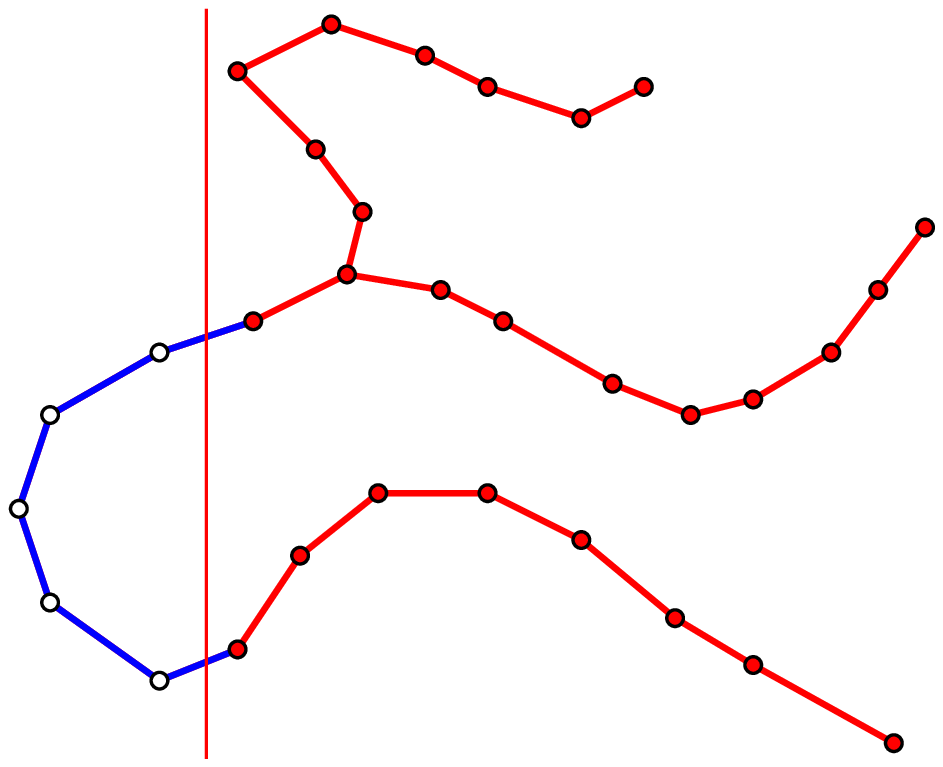}\hfill
\mbox{}
\end{center}
\caption{\label{fig:threshold}%
Four threshold sets for the function shown in Figure~\ref{fig:TDE},
with increasing threshold value from left to right.  The component
born at the far left only dies as it enters the far right, while the
much shorter-lived component born left of center dies entering the
very next step.}
\end{figure}

Observe that $\Gspace$ has a nested sequence of subgraphs, starting
with the empty subgraph,
\begin{equation}\label{eqn:filt}
  \emptyset
  =
  \Gspace(0)
  \subset
  \Gspace(1)
  \subset
  \Gspace(2)
  \subset
  \cdots
  \subset
  \Gspace(N)
  =
  \Gspace.
\end{equation}
New components appear and then join with older components as the
threshold parameter increases.  For the graph in Figure~\ref{fig:TDE},
four snapshots in this evolution appear in Figure~\ref{fig:threshold}.

If $\beta_0(i) = \beta_0(i-1)$, then $h(v_i)$ is a \emph{regular
value}; this happens precisely when $L(i)$ is a single vertex.
Otherwise, $h(v_i)$ is a \emph{critical value}.  In
Figure~\ref{fig:TDE}, the critical values are the $h$-values of the
letter-labeled vertices.

When $h(v_i)$ is a critical value, precisely one of the following two
things happens when passing from $\Gspace(i-1)$ to $\Gspace(i)$.
\begin{itemize}\itemsep=-0.2\baselineskip\vspace{-1ex}
\item%
$\beta_0(i) = \beta_0(i-1) + 1$: this happens when $L(i)$ is empty.
In this case, a new component $C_i$ is \emph{born} at $h(v_i)$, and we
associate~$C_i$ with $v_i$ for the rest of its existence.  The first
birth in our example happens at $h(A)$, where the threshold graph
changes from the empty set to a single vertex.  Subsequent component
births can be seen in the far left and center left of
Figure~\ref{fig:threshold}.
\item%
$\beta_0(i) = \beta_0(i-1) - k$ for some integer $k \geq 1$: this
happens when $L(i)$ consists of $k + 1$ vertices.  In this case, $k$
components \emph{die} at $h(v_i)$; the only one that remains alive is
the one associated to the vertex in $L(i)$ with lowest $h$-value.  For
example, referring again to Figure~\ref{fig:threshold}, the components
born at the far left and center left die when entering the far right
and center right, respectively.
\end{itemize}\vspace{-1ex}%

\subsubsection{Persistence diagrams}\label{sub:persistence-diagrams}

The evolution of connected components is compactly summarized in the
\emph{persistence diagram} $\Dgm_0(h)$.  This is a multiset of dots in
the plane~$\mathbb{R}^2$, meaning that each dot occurs with positive
integer (or infinite) multiplicity.  A dot of multiplicity $k$ at the
planar point $(a,b)$ indicates that $k$ components are born at
$h$-value~$a$ and die at $h$-value~$b$.  A component that is born at
$a$ but never dies corresponds to a dot $(a,\infty)$ in the diagram,
so $\mathbb{R}^2$ is extended to allow such points.  All of the dots
in~$\Dgm_0$ lie above the major diagonal $y = x$, since birth must
always precede death.  For technical reasons that will become clear
below, we also add a dot of infinite multiplicity at each point
$(x,x)$ on the major diagonal itself.  The diagram for our example is
on the right side of Figure \ref{fig:TDE}.

The \emph{persistence} of a dot $u = (a,b)$ is defined to be $pers(u)
= b - a$, the vertical distance to the major diagonal.  Each such dot
corresponds to a component $C$ that
\begin{itemize}\itemsep=-0.2\baselineskip\vspace{-1ex}
\item%
is not present in any of the graphs before threshold value~$a$,
\item%
exists as its own independent component for every threshold value
between $a$ and~$b$, and
\item%
joins with another component, born at or before $a$, exactly at the
threshold value $b$.
\end{itemize}\vspace{-1ex}%
Hence the persistence $b-a$ indicates the lifetime of this feature as
an independent component.  The actual geometric meaning of this
lifetime can vary.  For example, in Figures~\ref{fig:TDE}
and~\ref{fig:threshold}, the small-persistence dot $u = (h(D),h(E))$
points to the existence of a small ``wobble'' in the graph, as seen by
the height function $h$.  On the other hand, the large-persistence dot
$v = (h(C),h(F))$ reflects an arm of the tree that is very long, again
as measured in the vertical direction.

A general heuristic for interpreting these diagrams is to say that the
smaller the persistence of a dot is, the more likely its corresponding
feature is to be caused by measurement error or other noise.  A small
change in the values of $h$, for example, could remove the small
wobble that $u$ indicates.  This interpretation can be given more
rigor by the Stability Theorem described below.  That said, there is
no guarantee of persistence being correlated with importance, just
with reliability.  Indeed, one of the findings in Section
\ref{sec:Detailed} is that dots of not-particularly-high persistence
have the most distinguishing power in our specific application.

\subsubsection{Stability}\label{sub:stability}

\begin{figure}
\begin{center}
  \includegraphics[scale=0.3]{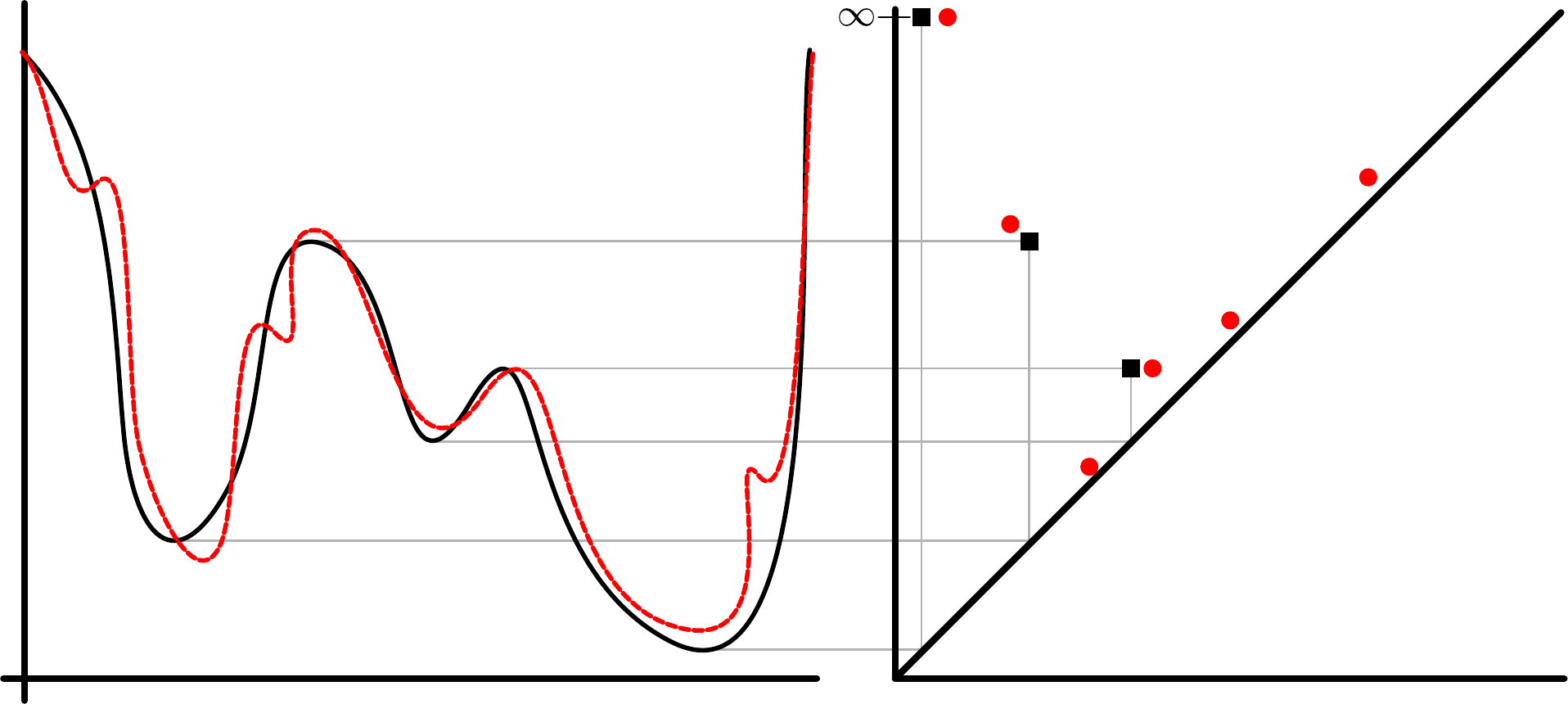}
\end{center}
\caption{\label{fig:noisy}%
On the left, functions $f$ and $g$, shown in black and red,
respectively.  On the right, their persistence diagrams, with the same
color scheme.  The optimal bijection between $\Dgm_0(g)$ and
$\Dgm_0(f)$ matches the three high-persistence dots together, and
matches the extra red low-persistence dots to the black diagonal.}
\end{figure}

We have outlined a tool that takes as input a real-valued function $h$
on a graph and returns a diagram $\Dgm_0(h)$ as output.  This tool
would be useless if it were not robust to small changes in the input
function.  For example, consider the left side of
Figure~\ref{fig:noisy}, which shows two functions $f$ (black) and $g$
(red) defined on the same interval; in this case, imagine that
$\Gspace$ is a simple path and the curves drawn indicate function
values as height.  Since $g$ is just a noisy version of $f$, data
analysis would fail to be robust if $\Dgm_0(g)$ differed too greatly
from $\Dgm_0(f)$.  Examining the right side of the same figure, the
major features of the two diagrams are very close, the only difference
being three extra red dots that lie very close to the diagonal,
corresponding to the low-persistence wobbles in the red function.

There are several metrics on the set of all persistence diagrams, but
all of them see these diagrams as being close to each other.
For example, let $D$ and $D'$ be two diagrams and choose a number $p
\in [1, \infty)$.  For each bijection $\phi: D \to D'$, define its
cost to be
$$%
  C_p(\phi) = \Big(\sum_{u \in D} ||u - \phi(u)||_p \Big)^{\frac1p}.
$$
Such bijections always exist, due to the infinite multiplicity of
every diagonal dot in each diagram.  The \emph{$p$-th Wasserstein
distance} $W_p(D,D')$ between the two diagrams is the infimum cost
$C_p(\phi)$ as $\phi$ ranges over all possible bijections.  Many
technical results (for example,
\cite{CohenSteiner2007, Cohen-Steiner2010}) all basically say that,
under mild conditions, $W_p(\Dgm_0(f),\Dgm_0(g)) \leq K ||f -
g||_{\infty}$.

\subsection{Thickening and loops}\label{sub:thickening-and-loops}

Having focused the previous discussion on connected components, we now
turn to the other type of persistence diagram used in our analysis.
Let $\Yspace$ be a compact subset of some Euclidean space $\Rspace^D$.
For each non-negative~$\alpha$, define $\Yspace_{\alpha}$ to be the
set of points in $\Rspace^D$ whose distance from $\Yspace$ is at
most~$\alpha$.  As $\alpha$ increases, loops appear and then
subsequently fill in.  Just as above, plot the birth and death times
of each loop as dots in the plane, and call the multiset of all such
dots the \emph{one-dimensional persistence diagram} $\Dgm_1(\Yspace)$.
\begin{figure}[hbt]
\centering
\hfill
\begin{tabular}[b]{@{}c@{}}
\includegraphics[width=.32\linewidth]{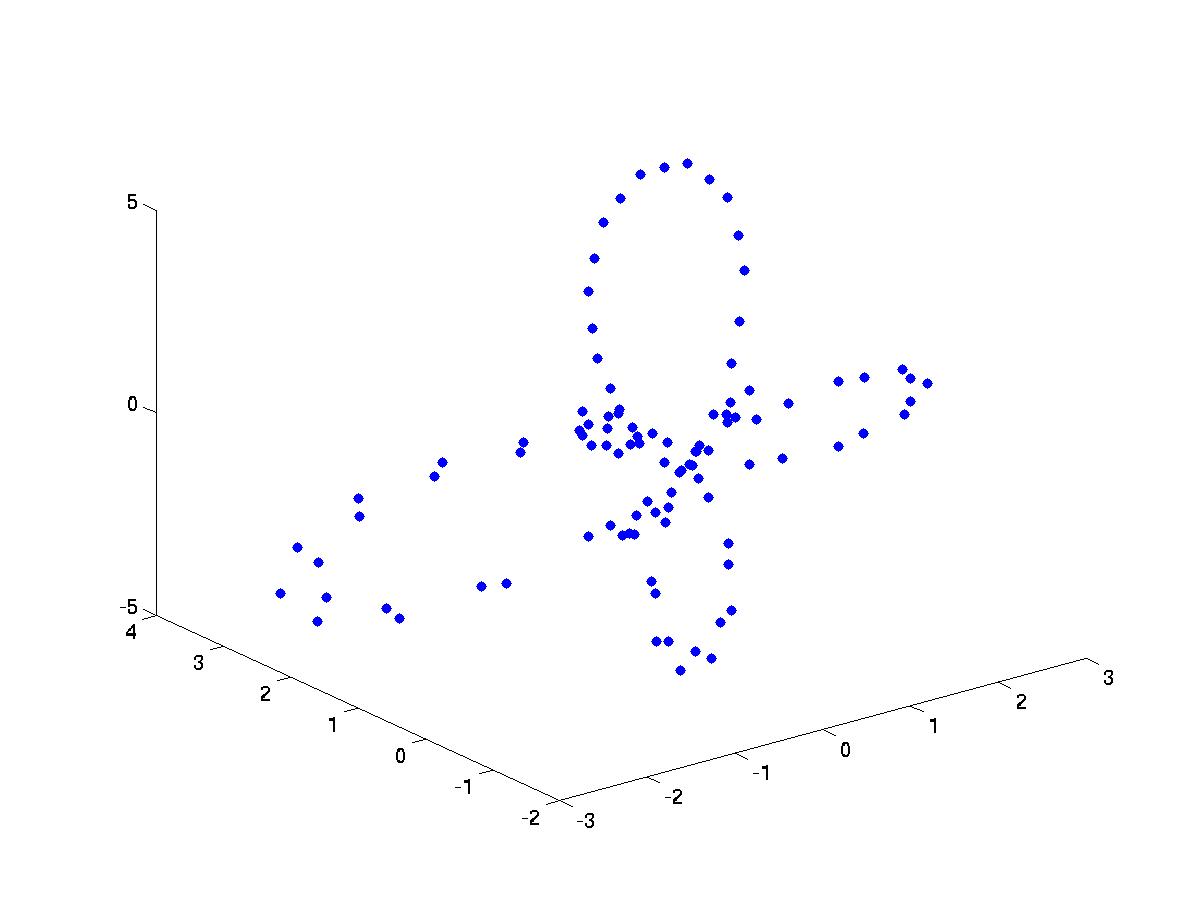}\\
(a) Point cloud $\Yspace$
\end{tabular}
\hfill
\begin{tabular}[b]{@{}c@{}}
\includegraphics[width=.32\linewidth]{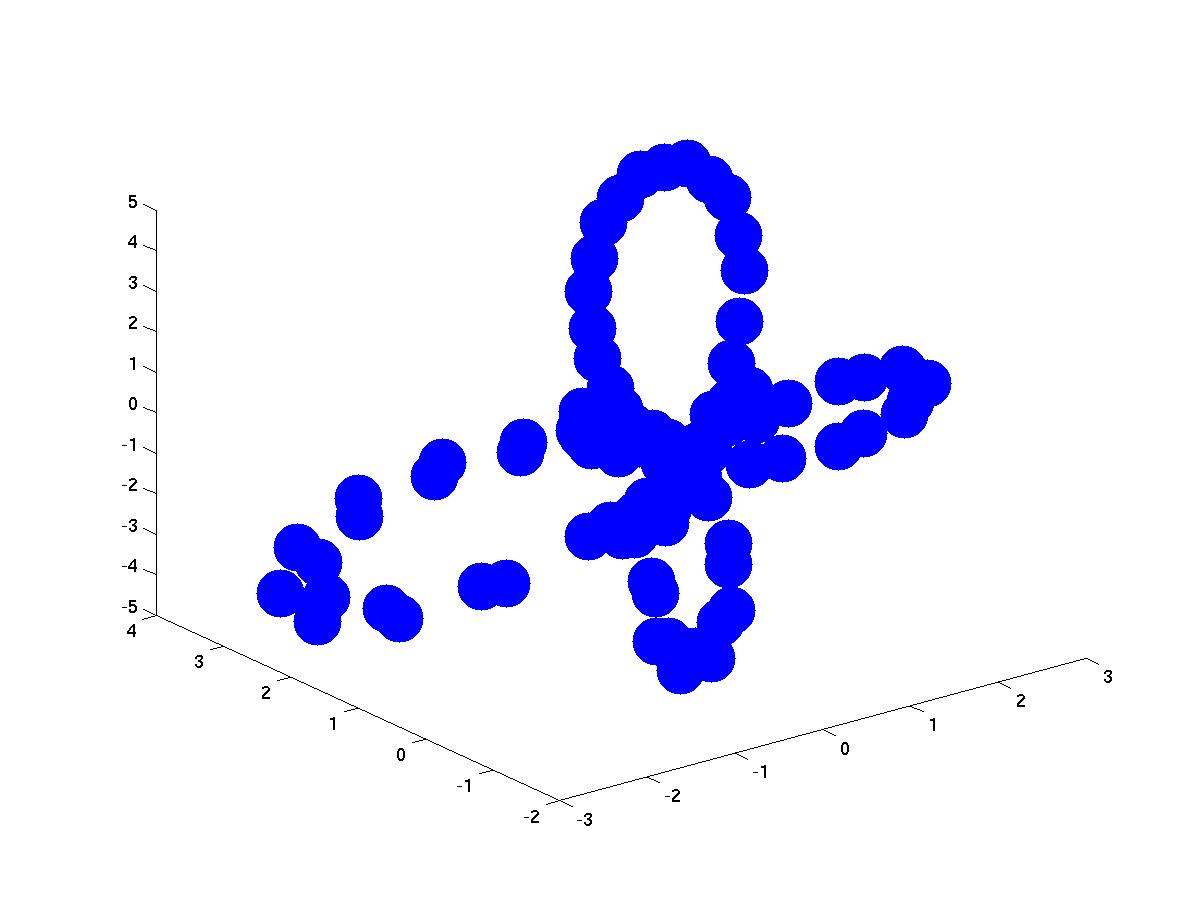}\\
(b) Thickening of $\Yspace$
\end{tabular}
\hfill
\begin{tabular}[b]{@{}c@{}}
\includegraphics[width=.32\linewidth]{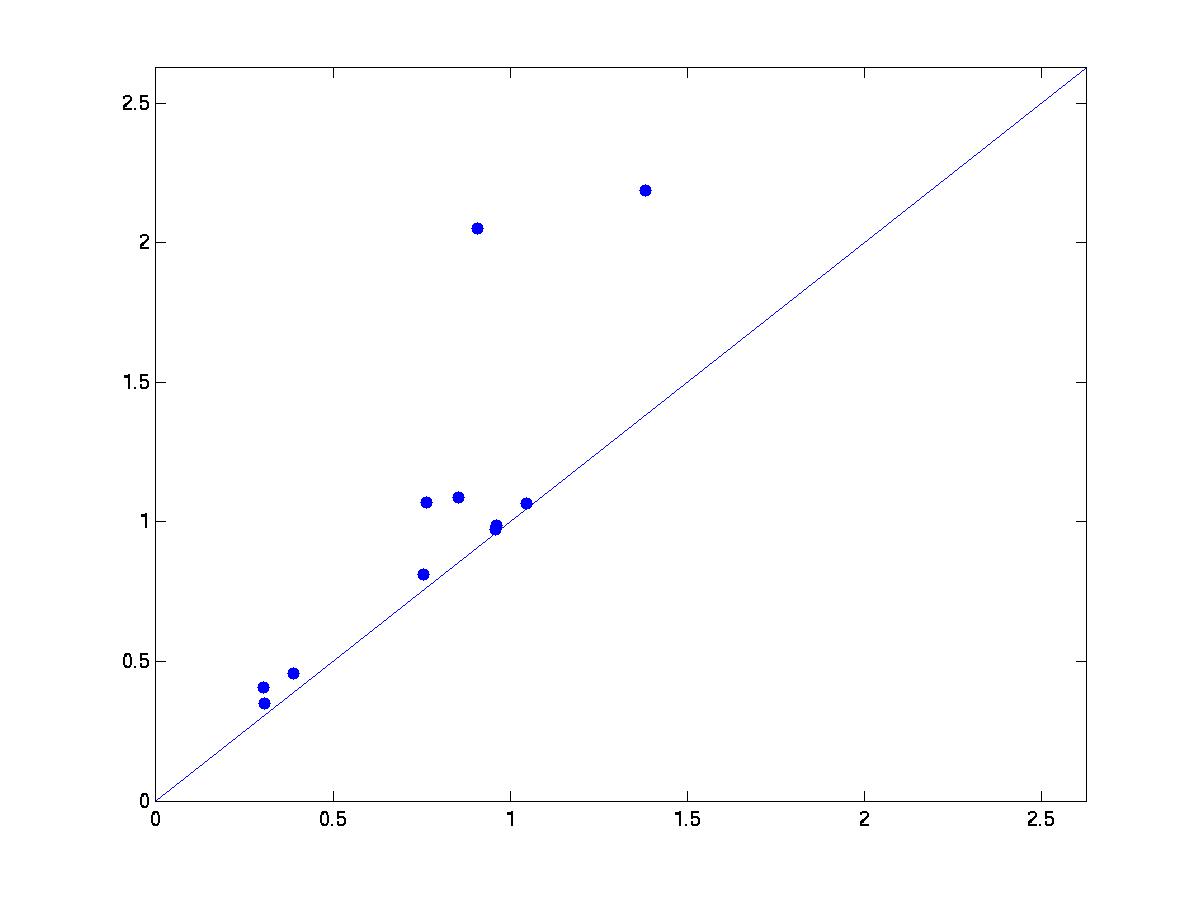}\\
(c) Persistence diagram $\Dgm_1(\Yspace)$
\end{tabular}
\hfill
\mbox{}
\caption{\label{fig:PC}%
Point cloud to persistence diagram}
\end{figure}

\subsubsection{Example}\label{sub:example}

Suppose that $\Yspace$ is the point cloud shown on the left of
Figure~\ref{fig:PC}.  Each $\Yspace_{\alpha}$ is the union of closed
balls of radius~$\alpha$ centered at the points of~$\Yspace$; one such
thickening is shown in the middle of the same figure.  The persistence
diagram $\Dgm_1(\Yspace)$ is on the right.

To explain the diagram, pretend that $\Yspace$ has been sampled from
an underlying space suggested by the point cloud.  The two dots of
highest persistence correspond to the two larger loops; the one with
later birth time corresponds to the leftmost loop, which reflects the
sparser density of sampling there.  The smaller loops are indicated by
the two dots of next highest persistence.  Finally, the group of dots
that sit almost on the diagonal are caused by little loops that
quickly come and go as the points thicken, as the result of holes
between small, increasingly overlapping convex sets; dots like these
would appear no matter what shape underlies the point cloud.

\subsubsection{Stability}\label{sub:stability'}

\begin{figure}[hbt]
\centering
\hfill
\begin{tabular}[b]{c}
\includegraphics[width=.33\linewidth]{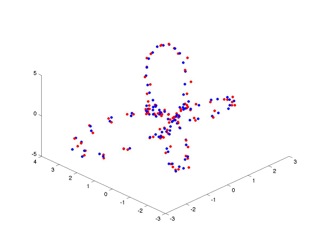}\\
(a) Point cloud $\Yspace'$
\end{tabular}
\hfill
\begin{tabular}[b]{c}
\includegraphics[width=.33\linewidth]{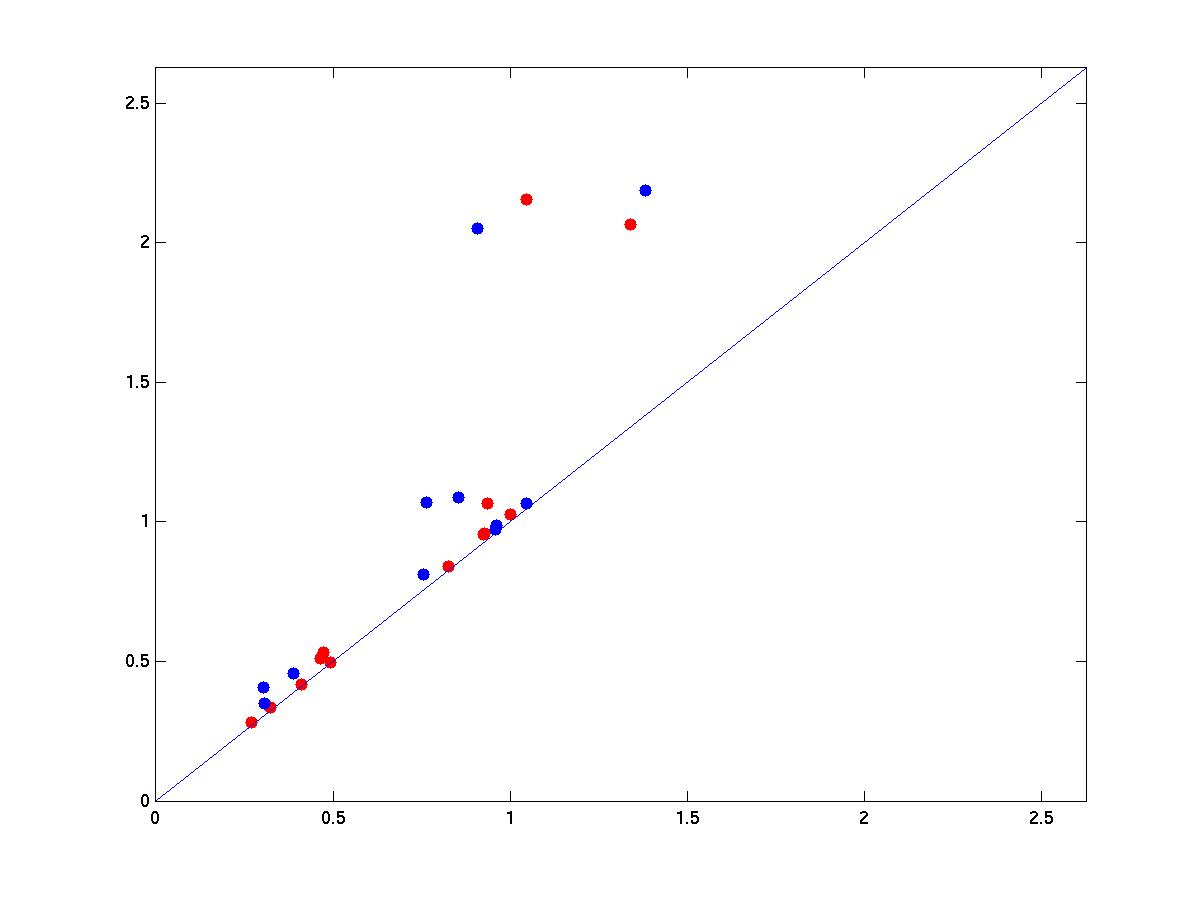}\\
(b) Persistence diagrams
\end{tabular}
\hfill
\mbox{}
\caption{\label{fig:NPC}%
Illustration of stability of persistence diagrams.  Two point clouds
(blue and red) that are close in Hausdorff distance are on the left,
and their corresponding persistence diagrams are on the right.\vspace{-3ex}}
\end{figure}

As with connected components, these loop diagrams are stable with
respect to perturbations of the input, in the following sense.  The
\emph{Hausdorff distance} $d_H(\Yspace, \Yspace')$ between two compact
sets is the smallest $\epsilon$ such that $\Yspace \subseteq
\Yspace'_{\epsilon}$ and $\Yspace' \subseteq \Yspace_{\epsilon}$.  For
example, the set $\Yspace'$ (in red) on the left of
Figure~\ref{fig:NPC} is close to our original set~$\Yspace$ (in blue)
in this metric.  The work of Cohen-Steiner et al.
(\cite{CohenSteiner2007}, \cite{Cohen-Steiner2010}) implies that
$W_p(\Dgm_1(\Yspace), \Dgm_1(\Yspace')) \leq K \cdot d_H(\Yspace,
\Yspace')$, as illustrated on the right of this figure.  A powerful
consequence of this result arises when $\Yspace'$ is a small but dense
sub-sampling of $\Yspace$: stability ensures that the persistence
diagram $\Dgm_1(\Yspace)$ can be well approximated by the diagram
derived from the sub-sample, a fact we apply in our analysis of brain
artery trees.

\subsection{From trees to diagrams}\label{sub:From-trees}

For persistence via connected components, our function~$h$ on each
tree $T$ is height: the value $h(v)$ at each vertex $v = (x,y,z)$ is
its third coordinate~$z$, and on each edge $(u,v)$ the value is
$h(u,v) = \max\{h(u),h(v)\}$.  We computed $\Dgm_0(h)$ as in
Section~\ref{sub:height}, with a simple (and fast) union-find
algorithm, running in $O(N\log{N})$, where $N$ is the number of
vertices of~$T$.

The running time for one-dimensional persistence is much slower, so we
did not compute the full-resolution persistence diagrams $\Dgm_1(T)$
associated to the thickening of each tree $T$ within the brain.
Instead, we sub-sampled each tree branch to produce a set of $3000$
total vertices per tree.  In contrast, each tree in the original
dataset has on the order of $10^5$ vertices, spread among roughly
200--300 tree branches.  The stability theorem for persistent homology
provides theoretical guarantees that our sub-sampling procedure does
not cause great change to the resulting persistence diagrams.

\begin{figure}[hbt]
\centering\vspace{-1ex}
\hfill
\begin{tabular}[b]{@{}c@{}}
\includegraphics[width=.29\linewidth]{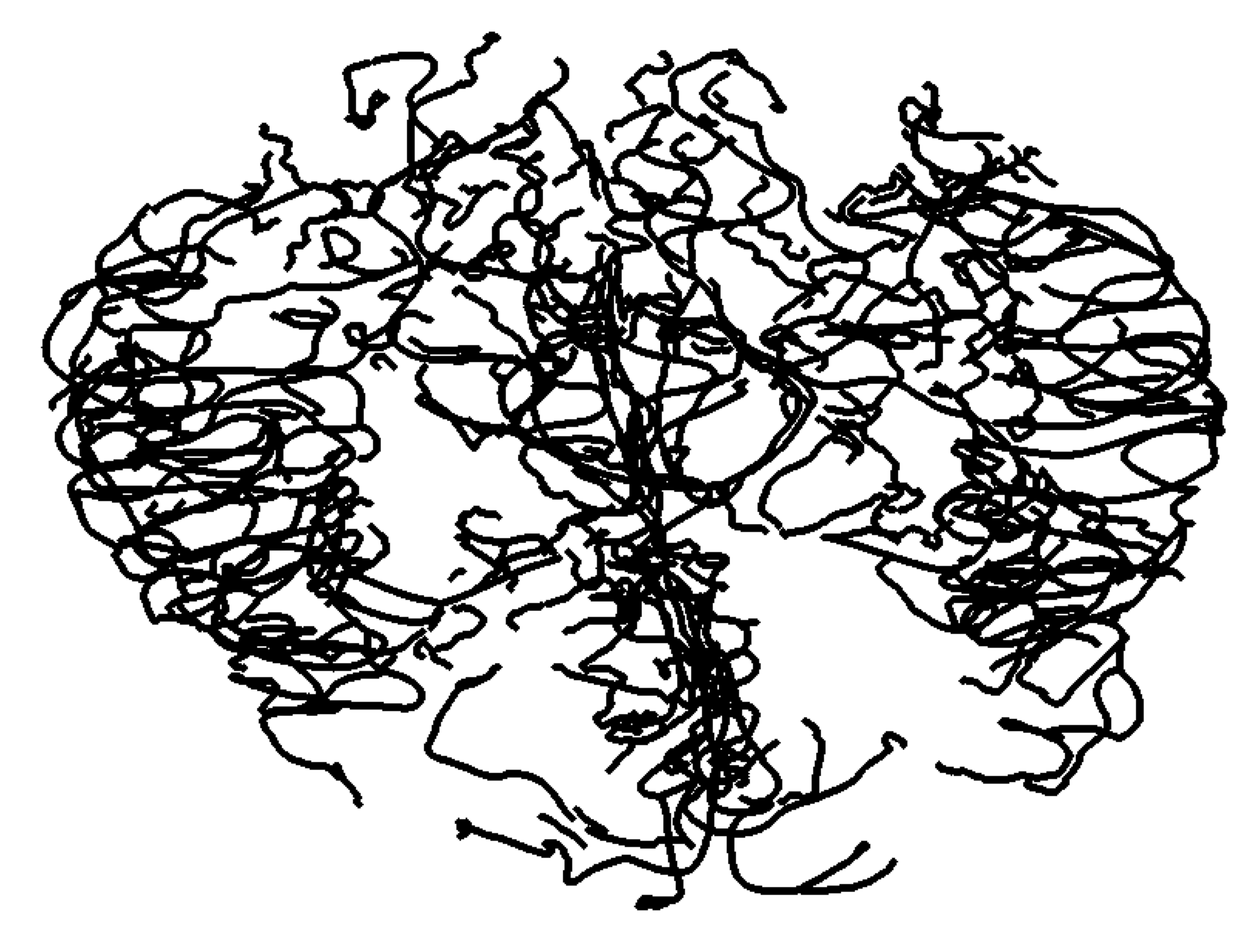}\\
(a) Brain tree
\end{tabular}
\hfill
\begin{tabular}[b]{@{}c@{}}
\includegraphics[width=.29\linewidth]{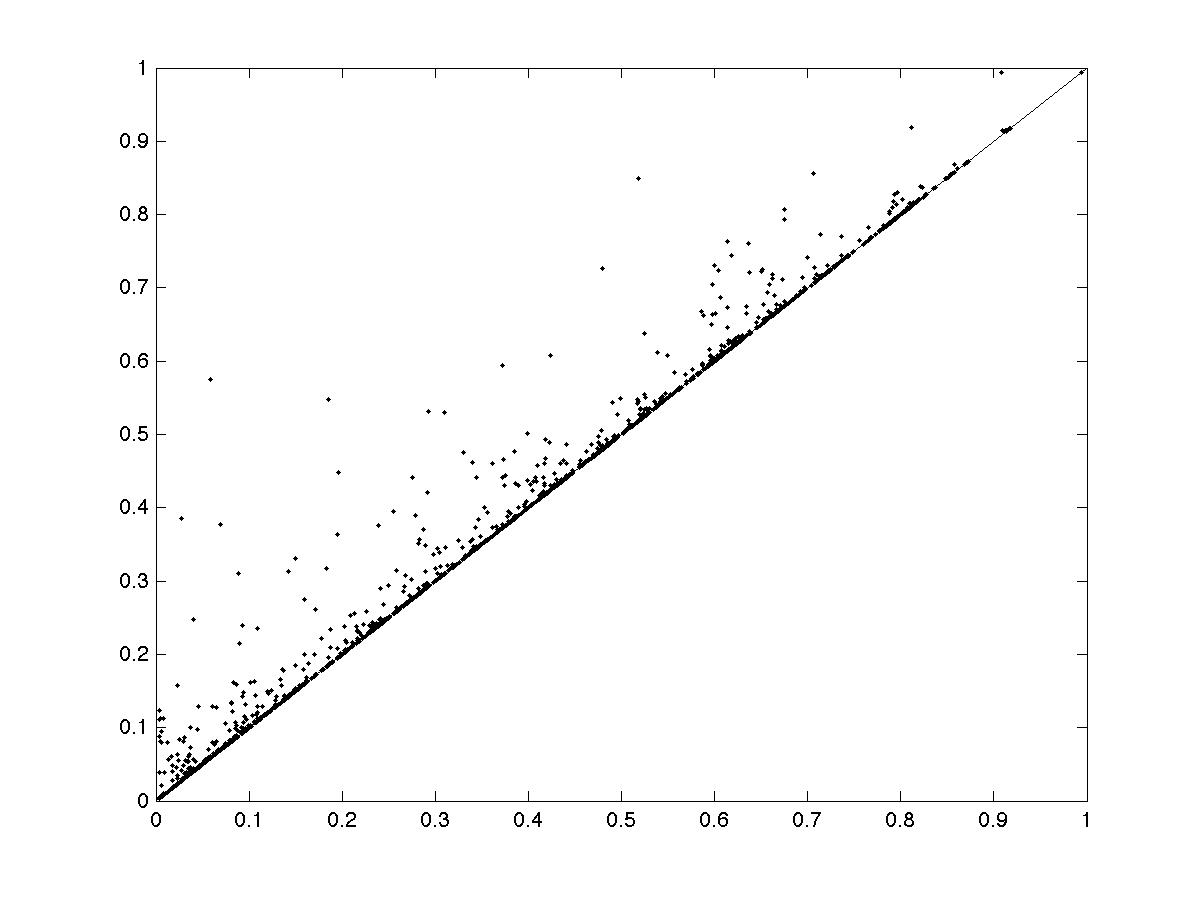}\\
(b) $\Dgm_0$
\end{tabular}
\hfill
\begin{tabular}[b]{@{}c@{}}
\includegraphics[width=.29\linewidth]{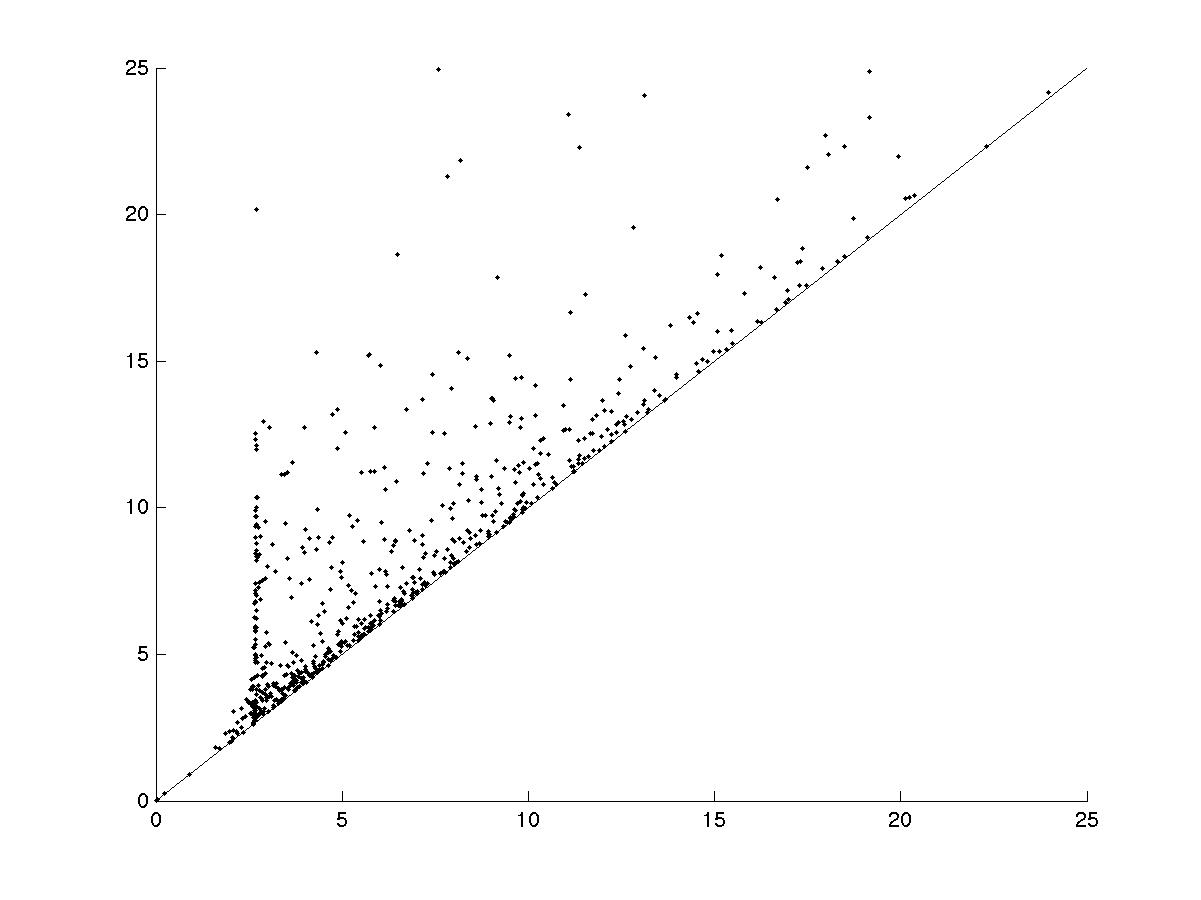}\\
(c) $\Dgm_1$
\end{tabular}\vspace{-1ex}
\hfill
\mbox{}
\caption{\label{fig:Age24}%
Persistent homology data objects from a 24-year old.  Left: brain
tree.  Middle: zero-dimensional diagram.  Right: one-dimensional
diagram.\vspace{-5ex}}
\end{figure}

\begin{figure}[hbt]
\centering\vspace{-1ex}
\hfill
\begin{tabular}[b]{@{}c@{}}
\includegraphics[width=.29\linewidth]{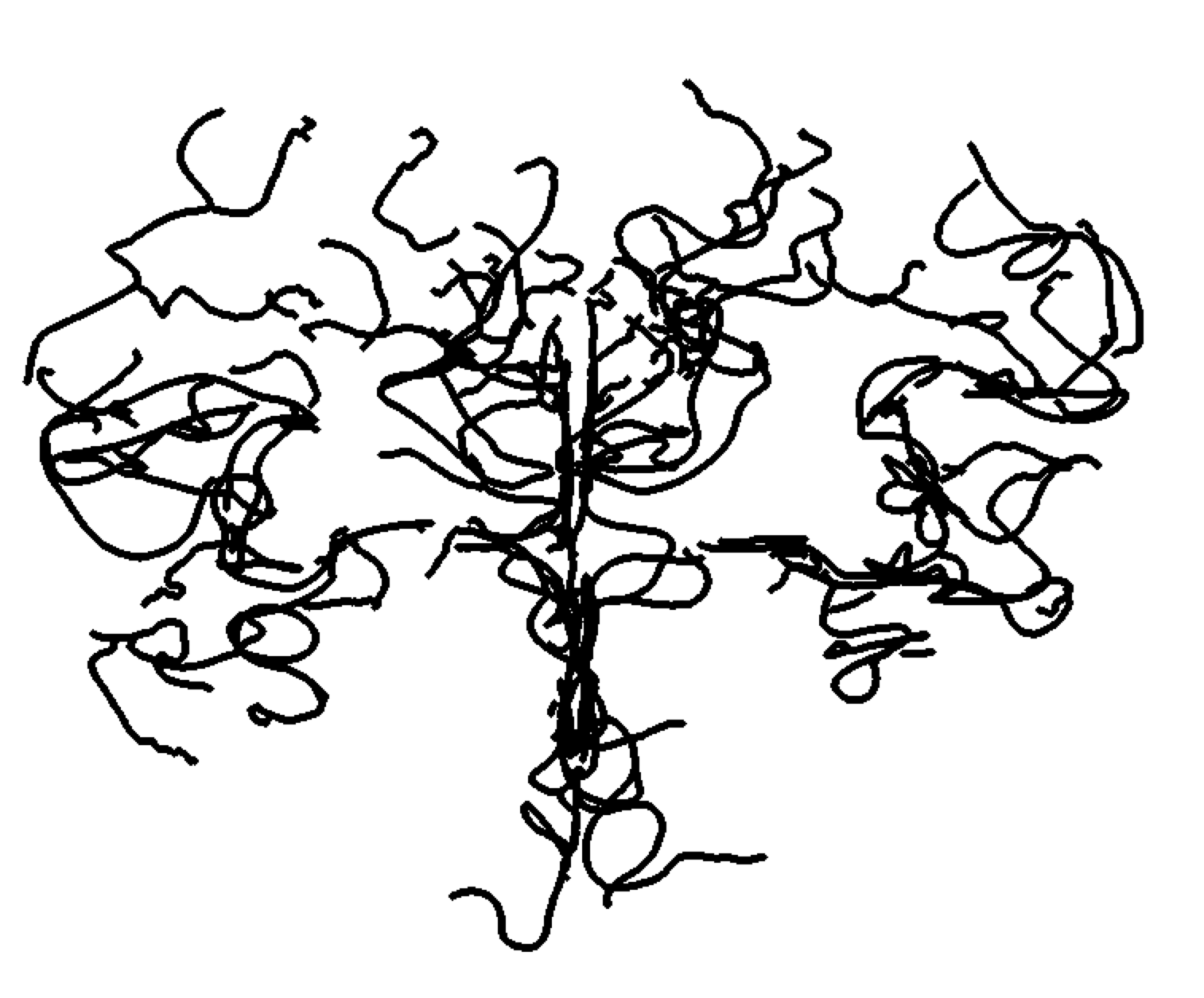}\\
(a) Brain tree
\end{tabular}
\hfill
\begin{tabular}[b]{@{}c@{}}
\includegraphics[width=.29\linewidth]{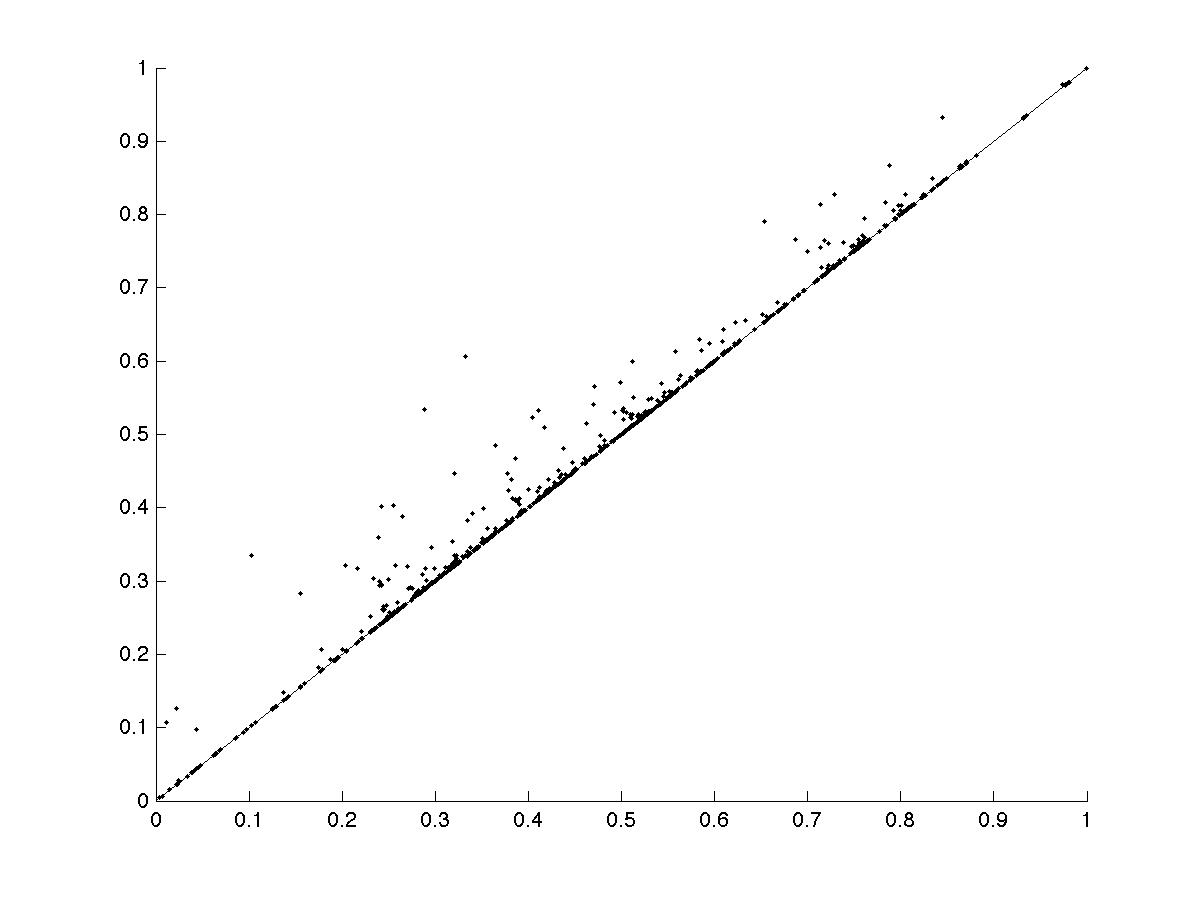}\\
(b) $\Dgm_0$
\end{tabular}
\hfill
\begin{tabular}[b]{@{}c@{}}
\includegraphics[width=.29\linewidth]{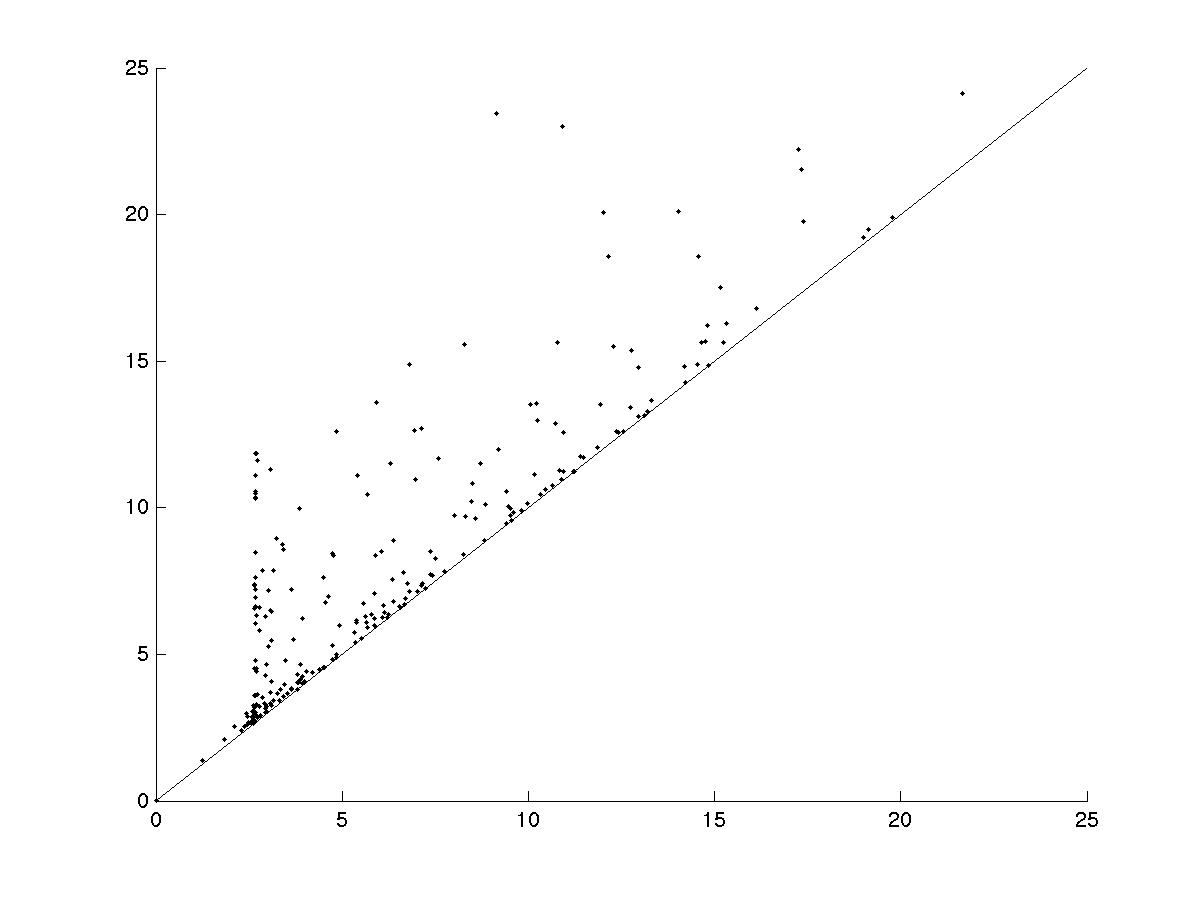}\\
(c) $\Dgm_1$
\end{tabular}\vspace{-1ex}
\hfill
\mbox{}
\caption{\label{fig:Age68}%
Persistent homology data objects from a 68-year old.  Left: brain
tree.  Middle: zero-dimensional diagram.  Right: one-dimensional
diagram.\vspace{-5ex}}
\end{figure}

Figure \ref{fig:Age24} shows the results of this analysis on the brain
tree of a $24$-year old subject: from left to right are the brain
tree, the $0$-dimensional diagram, and the $1$-dimensional diagram.
Compare this to Figure \ref{fig:Age68}, which shows a $68$-year old
subject.  Some qualitative differences might be noticed from these two
diagrams, but to give them any quantitative backing requires actual
statistical analysis of the diagram population, which we describe in
the next section.

\section{Detailed analysis of brain artery data}\label{sec:Detailed}

The methods in prior sections generate persistence diagrams to
summarize brain artery trees.  {}From there, statistical analysis can
proceed either with further summarization or without.  As the analysis
in Section~\ref{sec:PH} shows, vector-based summaries can capture
substantial structure while maintaining the possibility to apply the
full range of standard statistical analyses.  This section describes
our approach in more detail and then examines the effect of changes in
feature selection.

Our admittedly ad hoc method to turn diagrams into feature vectors is
justified somewhat by the nature of the geometry it is intended to
capture, but also by the excellent age and sex effects it reveals.
Other approaches to the same problem include binning the plane and
then turning each diagram into an integral vector consisting of bin
counts, as in \cite{Tracking2014}, or basing features
on more sophisticated algebraic geometry, as advocated in \cite{Adcock2013}.

We settled on vector-based analyses as a middle ground.  In general,
simple numerical summaries, such as total persistence or total number
of dots, miss too much useful information to be potent.  At the
opposite extreme, it is possible to work directly populations of
persistence diagrams, basing the analysis on metrics such as the
Wasserstein metric $W_p$ in Section~\ref{sub:stability}.  For example,
Gamble and Heo \cite{Gamble2010} found interesting structure using
multidimensional scaling with a $W_p$-dissimilarity matrix computed
from a set of persistence diagrams, each one associated with a set of
landmarks on a single tooth.  One could go further, using methods such
as the Fr\'echet mean approach of Mileyko et al.~\cite{Mileyko2011} to
find the center of the data followed by multidimensional scaling to
analyze variation about the mean.  We opted not to go that route
because computation of the $W_p$-metric is generally very expensive.
Another possibility, which we have not yet investigated, would be to
use Bubenik's theory of persistence landscapes \cite{bubenikLandscapes} to
translate the problem into one of functional data analysis.

\paragraph{Initial approach.}

For each of the $n = 98$ zero-dimensional persistence diagrams, we
computed the persistence of each dot; recall a dot has coordinates
$(b,d)$, where $b$ is birth and $d$ is death, and that its persistence
is $d - b$.  We then sorted these persistences in descending order,
and picked the first $100$ to produce a vector $(p_1, p_2, \ldots,
p_{100})$ for each brain.  In other words, the $i$-th coordinate of
this vector represents the size of the $i$-th largest ``bend'' in the
brain, as measured in the vertical direction.  The same procedure on
the one-dimensional diagrams led to the vectors $(q_1, q_2, \ldots,
q_{100})$, so that the number $q_j$ represents the size of the $j$-th
most persistent loop in the brain.  Both sets of vectors were used in
the age and sex analyses in Section~\ref{sec:PH}.

\paragraph{Feature scale.}

Are the observed age correlations being driven more by the
high-persistence features or by the lower-scale ones?  In addition,
does restricting to the $100$ most persistent dots miss useful
information?  To pursue these questions, we created the following sets
of vectors, for each pair of positive integers $n < N \leq 200:$
\begin{itemize}\itemsep=-0.2\baselineskip\vspace{-1ex}
\item%
$\mathbf{p}_{n,N} = (p_n, \ldots, p_N) \in \mathbb{R}^{N-n+1}$,
\item%
$\mathbf{q}_{n,N} = (q_n, \ldots, q_N) \in \mathbb{R}^{N-n+1}$.
\end{itemize}\vspace{-1ex}%
Using this notation, the original vectors used in our analysis were
$\mathbf{p}_{1,100}$ and $\mathbf{q}_{1,100}$.

\begin{figure}
\begin{center}
\includegraphics[scale=0.3]{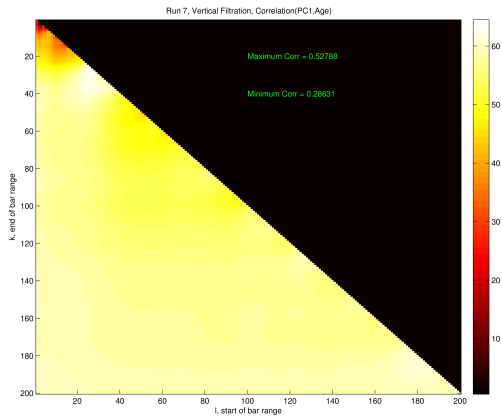}
\vspace{-3ex}
\end{center}
\caption{\label{fig:HM0}%
Age correlation heat map for features extracted from zero-dimensional
persistent homology analysis.  Color indicates the value of the
function $\rho(n,N)$, which is the age correlation derived from the
vectors $\mathbf{p}_{n,N}$, with $n$ on the horizontal axis and $N$ on
the vertical.  The upper-right black triangle is meaningless, as $n >
N$ does not lead to a vector in our scheme.}
\end{figure}

Extensive analysis of this feature set led to the heat map shown in
Figure~\ref{fig:HM0}.  The horizontal and vertical axes indicate $n$
and $N$, respectively, while the color at coordinates $(n,N)$ shows
the age correlation value $\rho(n,N)$ obtained by running our analysis
on the vectors $\mathbf{p}_{n,N}$.  Color in the lower diagonal part
of the plot codes correlation, ranging from very dark red (lowest)
through hotter colors to white (highest correlation).  The bottom of
the color range is $0.29$ and the top is $0.56$, chosen to maximize
use of the color scale.

Figure~\ref{fig:HM0} contains a lot of useful information.  First, the
small red area in the upper left indicates that the
highest-persistence features alone had far less distinguishing power
with respect to age; indeed, the two highest persistences $p_1$ and
$p_2$ lead only to an age correlation of $\rho = 0.26$.  On the other
hand, the rest of the lower triangle shows a fairly uniform, and high,
age correlation, leading to the surprising conclusion that one need
only include some of the more medium-scale persistence features to
obtain good age effects.  In fact, the length of the $28$-th longest
bar alone is a numerical feature that yields near-optimal correlation.
The same analysis performed on the one-dimensional features produced a
very similar pattern, not shown here.

The medium scale at which age correlation is optimized suggests a
reason why, in the initial stages of our connected component analysis
(Section~\ref{sub:height}), we found negligible differences in the
strength of correlation or significance upon filtering in various
directions other than upward.  Probably it is due to the stochastic
nature of blood vessel formation in the brain at the relevant scale:
while large features common to all human brains might have natural
ventral-dorsal orientation---such might be the case for major arteries
that branch from the circle of Willis and arch up to the top of the
brain and back down---the medium-sized features driving the observed
correlations are apparently random enough to be devoid of natural
orientation, statistically speaking.

\begin{figure}
\begin{center}
\includegraphics[scale=0.4]{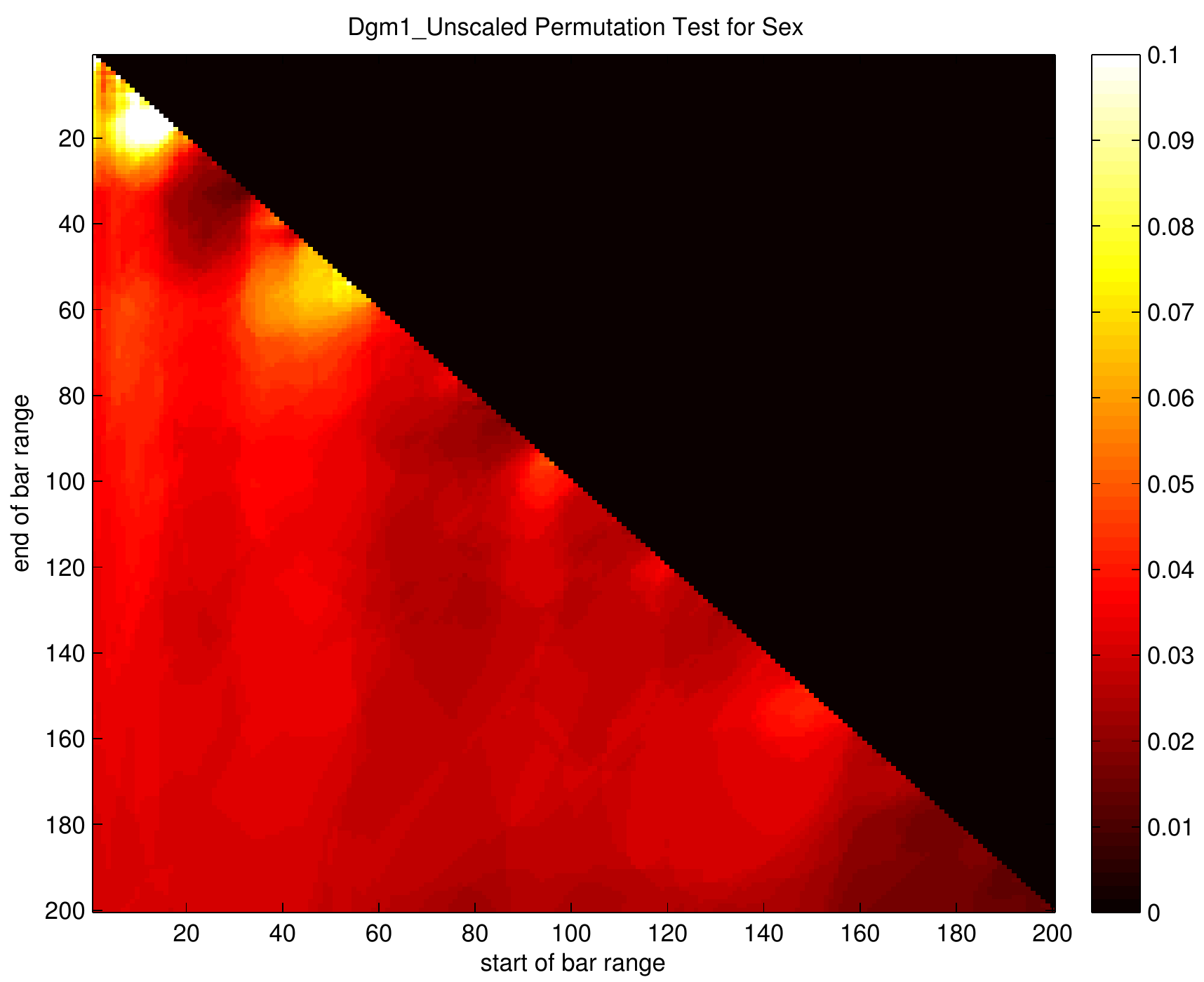}
\end{center}
\caption{\label{fig:HM1}%
Sex difference significance heat map for features extracted from
one-dimensional persistent homology analysis.  Color indicates the
value of the function $p(n,N)$, which is the significance, derived via
permuation test, of the difference between the male and the female
vectors $\mathbf{q}_{n,N}$, with $n$ on the horizontal axis and $N$ on
the vertical.  The upper-right black triangle is meaningless, as $n >
N$ does not lead to a vector in our scheme.  To provide good contrast
between the values, a color scheme running from $0.1$ (white) to $0$
(black) was chosen.  A few values are actually above $0.1$ and are
simply shown as white in this scheme.}
\end{figure}

Recall from Section~\ref{sec:PH} that a permutation test on the
vectors $\mathbf{q}_{1,100}$ found a significant ($p = 0.032$)
separation between male and female subjects.  One can also calculate
the sex-difference significance $p(n,N)$ obtained by running an
identical analysis on the vectors $\mathbf{q}_{n,N}$.  The resulting
pattern turns out to be similar to the findings for age correlation,
but even more stark.  Analyses that use only the most persistent loops
do not give clear sex separation; for example, $p(1,2)$ is only
$0.21$.  On the other hand, every single value of $p(n,N)$ with $N >
30$ lands below the significance level of $0.05$, with the minimum
value being $p(189,192) = 0.013$.  The heat map in Figure
\ref{fig:HM1} displays all of these values at once (darker is lower,
and hence more significant), and the near-uniformity of the
sex-difference significance is evident.

\section{Discusssion}\label{sec:Discussion}

This paper takes analysis of the brain artery tree data in the
entirely new direction of persistent homology.  This topological
analysis approach to data object representation gives stronger results
than those from alternative representations used in earlier studies.
This is the first study to find significant results even after
controlling for total artery length.

The lessons here are intended to suggest the power of these tools,
rather than to be anatomically conclusive, so multiple comparison
issues have not been carefully accounted for.  This serves to make the
main ideas more accessible than they would be with a careful
family-wise error rate, or false discovery rate analysis.  Interesting
future work is to apply these powerful new methods to other data sets
of tree-structured (or otherwise $1$-dimensional) objects.  An
important example of this is the airway data set of Feragan et
al. \cite{feragen2013toward}.

Finally, we note that the original data objects under consideration in
this paper were not the actual sets of arteries in $98$ human brains;
rather, they were the outputs of $98$ runs of the tube-tracking
algorithm from \cite{aydin2009principal}.  Like
all algorithms that process raw data, that algorithm introduces
artifacts, leading to the worry that analysis of its output data
objects may be picking up more on error than on signal.  In our case,
this worry applies to the zero-dimensional analysis, which looks at
component evolution in a given tree.  In contrast, the loop analysis
thickens a point sample from each tree into three-dimensional space,
so the stability theorem for persistent homology ensures that
replacing the given tree with a slightly modified version---even one
whose connectivity properties differ from the output of the
tube-tracking algorithm---does not cause great changes in the
persistence diagram.  An interesting new paper \cite{MolinaAbril2014} uses persistent homology methods to
aid in artifact-reduction in the actual ``upstream'' production of the
artery trees.  It would be valuable to run our analytical methods on
these new data objects to see if significant changes result.

\section*{Acknowledgments}

Support was provided by NSF grant DMS-1001437 for Miller and by the
Research Training Grant NSF-DMS 1045133 for Pieloch.  Ideas for thi
paper arose from discussions at the Mathematical Biosciences Institute
(MBI, DMS-0931642) and the Statistical and Applied Mathematical
Sciences Institute (SAMSI, DMS-1127914).  Pieloch thanks the
Information Initiative at Duke (iiD) for hosting him during Summer
2014.  The magnetic resonance brain images from healthy volunteers
used in this paper were collected and made available by the CASILab at
The University of North Carolina at Chapel Hill and were distributed
by the MIDAS Data Server at Kitware, Inc.


\end{document}